\documentclass[doublespacing,final,12pt]{elsart}
\usepackage{showkeys}
\usepackage{graphicx}
\journal{Nuclear Physics A}
\renewcommand{\listfigurename}{Figure Captions}

\begin{document}
\begin{frontmatter}
\title{ {Light Cone QCD Sum Rules Analysis of the Axial $N \rightarrow \Delta$ Transition Form Factors }}

\author{T. M. Aliev\thanksref{aliev}}
\ead{taliev@metu.edu.tr}
\author{K. Azizi\thanksref{kazem}}
\ead{kazizi@newton.physics.metu.edu.tr}
\author{A. Ozpineci\corauthref{cor}}
\ead{ozpineci@metu.edu.tr}
\thanks[aliev]{Permanent address: Institute of Physics, Baku, Azarbaijan}

\thanks[kazem]{kazizi@newton.physics.metu.edu.tr}

\corauth[cor]{Corresponding Author, tel: +90 312 210 3275, fax: +90 312 210 5099}
\address{Physics Department, Middle East Technical University, 06531, Ankara, Turkey}
 \date{}

\begin{abstract}
The axial $N-\Delta(1232)$ transition form factors are calculated within the light
cone QCD sum rules method. A comparison of our
results with the predictions of lattice theory  and quark model
is presented.
\end{abstract}
\end{frontmatter}
\setcounter{page}{1}
\section{Introduction}
Form factors have great importance in the investigation of the
internal structure of baryons. The inner structure of the nucleon is
encoded in several form factors. The electromagnetic form factors of
the nucleon, which parameterize the matrix element of the
electromagnetic current operators, are measured in a wide range of
$q^{2}$ values \cite{1,2,3}. The interest in nucleon form factor has
been renewed by recent progress in experimental physics,
where it became possible to get polarized beams and polarized targets. This
possibility opened up two new methods \cite{4} to measure the ratio
of the electric and magnetic form factors, namely polarization
transfer and beam-target asymmetry. New experimental data  obtained
from $e+p\rightarrow e+p$ reaction, which is performed at JLAB \cite{5},
gave the result that the ratio $\mu_{\rho}G_{E}/G_{M}$ decreases
from unity substantially for high $Q^{2}$ values. Not only form
factors relevant for the diagonal transitions, but also the
electromagnetic transition form factors for electro-production of
the $\Delta$ has also been the subject of recent experimental
\cite{5,6,7} and theoretical studies \cite{8}.

The main advantage of the nucleon to $\Delta$ transition is that,
the $\Delta$ is a dominant nucleon resonance and its identification is
easy since the spin parity selection rules provide us information
about the wave function of these baryons. The $N-\Delta$ transition
form factors due to the weak axial current can give valuable
information about the structure of the baryons, complementary to
that obtained from electromagnetic transition. For example,
measurement of axial form factors for $N-\Delta$ transition, allows
us to check off diagonal Goldberger-Treiman relation, order of
conservation of the axial current, etc. Weak axial $N-\Delta$
transition form factors are investigated in neutrino (or charged
lepton) scattering on deuterium or hydrogen in the $\Delta$ region
experiments. New information on the weak axial form factors is
expected from parity-violating electron scattering experiments
planned at Jefferson Laboratory \cite{10}.

In the present work, we calculate the axial $N \to \Delta$ transition
form factors in light cone QCD sum rules (LCQSR). In the light cone
QCD sum rules method, operator product expansion is carried out near the
light cone $x^{2}\simeq0$, and non perturbative dynamics is
parameterized by the light cone distribution amplitudes that determine the matrix
elements of the nonlocal operators between vacuum and one particle
states. The expansion near the light cone is an expansion in the twist of the operators
rather than the
dimension as in the traditional QCD sum rules (for more about this method see e.g.
\cite{11,12} and references therein). The light cone distribution amplitudes of the proton is calculated in \cite{17,18,18p}.
Using the light cone distribution amplitudes of the proton semileptonic $\Lambda_b \rightarrow p \ell \nu$ \cite{16y} decay, scalar form factor of the proton \cite{17y,18y},
axial and induced pseudo scalar form factors of the nucleon \cite{19y, 20y}, $\Sigma-n$ form factors \cite{21y} are studied. In \cite{22y}, the distribution amplitudes of
$\Lambda$ in leading conformal spin is calculated and the obtained amplitudes are used to study the $\Lambda_c \rightarrow \Lambda \ell \nu$ decay. Note that these form factors are calculated in
lattice QCD in \cite{13,yeni}, and in chiral constituent quark model in
\cite{14}.

The plan of this work is as follows: In section 2 we consider the generic
correlator function and present the LCQSR formalism. In this section
we obtain the LCQSR for the axial $N$ to $\Delta$ transition form
factors. The numerical analysis and discussion is presented in
section 3.
\section{Sum Rules for the Axial Nucleon to Delta Transition Form Factors  }
In the present section, light cone QCD sum rules for the axial
$N$ to $\Delta$ transition form factors are derived. These form factors are
defined by the matrix element of the axial current (in our case
isovector part of the axial current), i.e.
\begin{equation}\label{axial}
J_{\nu}^{3}(x)=\overline{\psi}(x)\gamma_{\nu}\gamma_{5}\frac{\tau^{3}}{2}\psi(x)
\end{equation}
between the nucleon state with momentum p and the $\Delta$-state
with momentum $p'=p-q$, where $\tau^{3}$ is the third Pauli matrix,
and $q$ is the transferred momentum. Hence, the axial $N \to
\Delta$ transition form factors are defined by the matrix
element $<\Delta(p',s')\mid J_{\nu}^{3}\mid N(p,s)>$.
This $N$ to $\Delta$ weak matrix element can be expressed in terms
of four invariant transition form factors as follows \cite{15,16}:
\begin{eqnarray}\label{mat.el2}
<\Delta(p',s')\mid J_{\nu}^{3}\mid N(p,s)>&=&
i\overline{u}^{\lambda}(p',s')\{(\frac{C_{3}^{A}(q^{2})}{m_{N}}\gamma_{\mu}+\frac{C_{4}^{A}(q^{2})}{m_{N}^{2}}p'_{\mu})(g_{\lambda\nu}
g_{\rho\mu}\nonumber\\
&-&g_{\lambda\rho} g_{\mu\nu})q^{\rho}
+C_{5}^{A}(q^{2})g_{\lambda\nu}+\frac{C_{6}^{A}(q^{2})}{m_{N}^{2}}q_{\lambda}q_{\nu}\}u(p,s)\nonumber\\
\end{eqnarray}
where $u^{\lambda}(p',s')$ is the Rarita-Schwinger spinor for $\Delta$
and $u(p,s)$ is the nucleon spinor.
Note that partial conservation of axial current (PCAC) and pion meson
dominance leads to the relation \cite{yeni}
\begin{eqnarray}
C_6^A(Q^2) = \frac{m_N^2}{Q^2+m_\pi^2} C_5(Q^2)
\label{reln}
\end{eqnarray}

For the calculation of the above mentioned form factors within the
light cone QCD sum rule method we consider the matrix element in
which one of hadrons is described by an interpolating current with
quantum numbers of this hadron and another one is represented by the
state vector. For the construction of the light cone sum rules, the
distribution amplitudes (DA's) of state vector hadron is needed. The
nucleon distribution amplitudes for all three quark operators are
calculated in \cite{17,18,18p} and DA's for $\Delta$ isobar is not yet
calculated. For this reason, we consider the following correlator
function where nucleon is described by state vector
\begin{equation}\label{T}
\Pi_{\mu\nu}(p,q)=i\int d^{4}xe^{iqx}<0\mid
T\{\eta_{\mu}(0)J^{3}_{\nu}(x) \}\mid N(p)>
\end{equation}
The interpolating current for the $\Delta^{+}$ isobar is chosen in
the form \cite{19}
 \begin{equation}\label{cur.Del}
\eta_{\mu}(0)=\frac{1}{\sqrt{3}}\varepsilon^{abc}[2(u^{aT}(0)C\gamma_{\mu}d^{b}(0))u^{c}(0)+(u^{aT}(0)C\gamma_{\mu}u^{b}(0))d^{c}(0)]
\end{equation}
 and the axial  current is:
\begin{equation}\label{cur.Axial}
J^{3}_{\nu}(x)=\frac{1}{2}[\overline{u}(x)\gamma_{\nu}\gamma_{5}u(x)-\overline{d}(x)\gamma_{\nu}\gamma_{5}d(x)]\nonumber\\
\end{equation}

In order to construct the sum rules for the form factors,
 the correlation function will be represented in two different forms, i.e. in terms
of hadron parameters and in terms of the quark-gluon parameters. Let us first consider
the representation of the correlation function in terms of hadron parameters
(phenomenological part). Phenomenological part of the correlation
function can be obtained by inserting the complete set of hadrons
with the same quantum numbers of $\eta_{\mu}(0)$ inside the correlation function. Saturating the
correlator function with these hadrons and isolating the
contribution coming from ground state hadron we get
\begin{equation}\label{corr.func}
\Pi_{\mu\nu}(p,q)=\sum_{s'}\frac{<0\mid
\eta_{\mu}\mid\Delta^{+}(p',s')><\Delta^{+}(p',s')\mid
J_{\nu}\mid N(p,s)>}{m_{\Delta}^{2}-p'^{2}}+ \cdots
\end{equation}
where $m_{\Delta}$ is $\Delta^{+}$ mass and $\cdots$ represents
contributions from higher states and the continuum
. The matrix element
$<\Delta^{+}\mid J_{\nu}(x)\mid N>$ entering in Eq.
(\ref{corr.func}) is given by Eq. (\ref{mat.el2}) and the matrix
element $<0\mid \eta_{\mu}(0)\mid \Delta^{+}>$ is determined as
\begin{equation}\label{lambda}
<0\mid \eta_{\mu}(0)\mid \Delta^{+}>=\lambda_{\Delta}u^{\mu}(p',s')
\end{equation}
where $\lambda_{\Delta}$ is the residue of  $ \Delta^{+}$ isobar and
the value of the residue $\lambda_{\Delta}$ is determined from the
two point sum rules \cite{19,20,21,22,23}. Performing summation over
spin of Rarita-Schwinger spinor using the formula
\begin{equation}\label{spinor}
u_{\mu}(p',s)\overline{u_{\nu}}(p',s)=-(\not\!p'+m_{\Delta})\{g_{\mu\nu}-\frac{1}{3}\gamma_{\mu}\gamma_{\nu}-\frac{2p'_{\mu}p'_{\nu}}
{3m_{\Delta}^{2}}+\frac{p'_{\mu}\gamma_{\nu}-p'_{\nu}\gamma_{\mu}}{3m_{\Delta}}\}
\end{equation}
and using Eqs. (\ref{mat.el2}), (\ref{corr.func}), (\ref{lambda})
and (\ref{spinor}) the contribution of $\Delta^{+}$ to the
correlation function can be written as
\begin{eqnarray}\label{phen.1}
\Pi_{\mu\nu}(p,q)&=&\frac{-i\lambda_{\Delta}}{m_{\Delta}^{2}-p'^{2}}(\not\!p'+m_{\Delta})\left[g_{\mu\lambda}-\frac{1}{3}\gamma_{\mu}\gamma_{\lambda}
-\frac{2p'_{\mu}p'_{\lambda}}
{3m_{\Delta}^{2}}+\left.\frac{p'_{\mu}\gamma_{\lambda}-p'_{\lambda}\gamma_{\mu}}{3m_{\Delta}}\right]\right\{\nonumber\\
&&\left[\frac{C_{3}^{A}(q^{2})}{m_{N}}\gamma_{\alpha}+\frac{C_{4}^{A}(q^{2})}{m_{N}^{2}}p'_{\alpha}\right](g_{\lambda\nu}q_{\alpha}-q_{\lambda}g_{\nu\alpha}
)
+C_{5}^{A}(q^{2})g_{\lambda\nu}\nonumber\\
&+&\left.\frac{C_{6}^{A}(q^{2})}{m_{N}^{2}}q_{\lambda}q_{\nu}\right\}u(p)
\end{eqnarray}
From Eq. (\ref{phen.1}) it follows that the correlator function
contains numerous tensor structures and not all of them are independent.
The dependence can be removed by ordering gamma matrices in a
specific order. In this work the ordering $\gamma_\mu
\not\!p'\gamma_{\nu}\not\!q$ is chosen. With this ordering the
correlation function becomes
\begin{eqnarray}\label{phen.2}
\Pi_{\mu\nu}(p,q)&=&\frac{-i\lambda_{\Delta}}{m_{\Delta}^{2}-p'^{2}}\left\{g_{\mu\nu}\not\!qC_{3}^{A}(q^{2})+g_{\mu\nu}\not\!p'\not\!q
\frac{C_{3}^{A}(q^{2})}{m_{N}}\right.+g_{\mu\nu}(\not\!p'+m_{\Delta})\nonumber\\
&\times&\left(C_{5}^{A}(q^{2})+C_{4}^{A}(q^{2})\frac{p'q}{m_{N}^{2}}\right)-q_{\mu}(\not\!p'+m_{\Delta})\gamma_{\nu}\frac{C_{3}^{A}(q^{2})}{m_{N}}
\nonumber \\
&-&q_{\mu}(\not\!p'+m_{\Delta})p'_{\nu}\frac{C_{4}^{A}(q^{2})}{m_{N}^{2}}
+q_{\mu}q_{\nu}(\not\!p'+\left.m_{\Delta})\frac{C_{6}^{A}(q^{2})}{m_{N}^{2}}\right\}u(p)\nonumber\\
&+& \mbox{other structures with }\gamma_{\mu} \mbox{ at the beginning or
which are}\nonumber\\
 &&\mbox{proportional to }p'_{\mu}.
\end{eqnarray}
The reason why  the structures $\sim p'_{\mu}$ and structures with
$\gamma_{\mu}$ at the beginning is not considered is follows.
The interpolating current $\eta_{\mu}$ couples not only to 
spin-parity $(3/2)^{+}$ states, but also to 
spin-parity $(1/2)^{-}$ states. In other words $\eta_{\mu}$ has nonzero
overlap with spin $1/2$ states. This coupling can be written in the most general form as:
\begin{equation}\label{overlap}
<0\mid \eta_{\mu}\mid\frac{1}{2}(p')>=(Ap'_{\mu}+B\gamma_{\mu})u(p)
\end{equation}
Using the condition $\gamma^{\mu}\eta_{\mu}=0$ one can easily obtain
that $B=\frac{-Am_{1/2}}{4}$. Therefore, in general case spin $1/2$
states give also contribution to the considered correlation function
but they contribute only to the structures which contain a
$\gamma_{\mu}$ at the beginning or which are proportional to $p'_{\mu}$.
By choosing the ordering
$\gamma_{\mu}\not\!p'\gamma_{\nu}\not\!q$ and ignoring the structures proportional to $p'_\mu$ and the structures
that contain a $\gamma_\mu$ at the beginning, the contributions to the correlation function from states that have spin $1/2$
are eliminated.


Now we turn our attention to the calculation of the correlator
function from the QCD side for the large Euclidean virtuality
$p'^{2}=(p-q)^{2} << 0$ in terms of the nucleon distribution
amplitudes. Using the explicit expression for the $\Delta^{+}$
isobar interpolating current Eq. (\ref{cur.Del}) and axial current
Eq. (\ref{cur.Axial}), and carrying out all contractions
for the correlation function, Eq. (\ref{T}), in $x$ representation we get the
following result
\begin{eqnarray}\label{mut.m}
\Pi_{\mu\nu}(p,q)&=&\frac{-1}{16\pi^{2}\sqrt{3}}\int
\frac{d^{4}xe^{iqx}}{x^{4}}\{(C\gamma_{\mu})_{\alpha\beta}(\gamma_{\nu}\gamma_{5})_{\rho\sigma}\varepsilon^{abc}<0\mid[
4u_{\eta}^{a}(0)u_{\theta}^{b}(x )d_{\phi}^{c}(0)\nonumber\\
&&(2g_{\alpha\eta}g_{\sigma\theta} g_{\beta\phi}(\not\!x
)_{\lambda\rho}
+2g_{\lambda\eta}g_{\sigma\theta}g_{\beta\phi}(\not\!x
)_{\alpha\rho}+g_{\alpha\eta}g_{\sigma\theta}g_{\lambda\phi}(\not\!x
)_{\beta\rho}
+g_{\beta\eta}g_{\sigma\theta}g_{\lambda\phi}(\not\!x)_{\alpha\rho})
\nonumber \\
&-&4u_{\eta}^{a}(0)u_{\theta}^{b}(0)d_{\phi}^{c}(x
)(2g_{\alpha\eta}g_{\lambda\theta} g_{\sigma\phi}(\not\!x
)_{\beta\rho}+\left. g_{\alpha\eta}g_{\beta\theta}g_{\sigma\phi}(\not\!x
)_{\lambda\rho})]\mid N(p)>\right\}
\end{eqnarray}
where we have used the light cone expanded light quark propagator
as \cite{22}:
\begin{eqnarray}\label{prop}
S(x) &=& \frac{i\not\!x}{2\pi^2x^4}
-<qq>(1+\frac{m_{0}^{2}x^{2}}{16})-ig_{s} \int\limits_0^1dv
[\frac{\not\!x}{16\pi^2x^2}G_{\mu\nu}\sigma^{\mu\nu}\nonumber\\
&-&vx^{\mu}G_{\mu\nu}\gamma^{\nu}\frac{i}{4\pi^2x^2}]
\end{eqnarray}
The terms proportional to the gluon strength tensor can give
contribution to four and five particle distribution functions but
they are expected to be small  \cite{17, 18,18p} and for this reason we
will neglect these amplitudes in further analysis. The terms
proportional to $<qq>$ can also be omitted because Borel
transformation eliminates these terms and hence only first term in
Eq. (\ref{prop}) is relevant for our discussion. From Eq.
(\ref{mut.m}) it follows that for the calculation of
$\Pi_{\mu\nu}(p,q)$ we need to know the matrix element $$<0\mid[
4u_{\eta}^{a}(0)u_{\theta}^{b}(x)d_{\phi}^{c}(0)\mid N(p)>.$$ This
three quark matrix element between vacuum and the proton state is
given in \cite{17,18,18p} as:
\begin{eqnarray}\label{wave func}
&&4\langle0|\epsilon^{abc}u_\alpha^a(a_1 x)u_\beta^b(a_2
x)d_\gamma^c(a_3 x)|P\rangle
=\mathcal{S}_1m_{N}C_{\alpha\beta}(\gamma_5N)_{\gamma}+
\mathcal{S}_2m_{N}^2C_{\alpha\beta}(\rlap/x\gamma_5N)_{\gamma}\nonumber\\
&+& \mathcal{P}_1m_{N}(\gamma_5C)_{\alpha\beta}N_{\gamma}+
\mathcal{P}_2m_{N}^2(\gamma_5C)_{\alpha\beta}(\rlap/xN)_{\gamma}+
(\mathcal{V}_1+\frac{x^2m_{N}^2}{4}\mathcal{V}_1^M)(\rlap/pC)_{\alpha\beta}(\gamma_5N)_{\gamma}
\nonumber\\&+&
\mathcal{V}_2m_{N}(\rlap/pC)_{\alpha\beta}(\rlap/x\gamma_5N)_{\gamma}+
\mathcal{V}_3m_{N}(\gamma_\mu
C)_{\alpha\beta}(\gamma^\mu\gamma_5N)_{\gamma}+
\mathcal{V}_4m_{N}^2(\rlap/xC)_{\alpha\beta}(\gamma_5N)_{\gamma}\nonumber\\&+&
\mathcal{V}_5m_{N}^2(\gamma_\mu
C)_{\alpha\beta}(i\sigma^{\mu\nu}x_\nu\gamma_5N)_{\gamma} +
\mathcal{V}_6m_{N}^3(\rlap/xC)_{\alpha\beta}(\rlap/x\gamma_5N)_{\gamma}
+(\mathcal{A}_1\nonumber\\
&+&\frac{x^2m_{N}^2}{4}\mathcal{A}_1^M)(\rlap/p\gamma_5
C)_{\alpha\beta}N_{\gamma}+
\mathcal{A}_2m_{N}(\rlap/p\gamma_5C)_{\alpha\beta}(\rlap/xN)_{\gamma}+
\mathcal{A}_3m_{N}(\gamma_\mu\gamma_5 C)_{\alpha\beta}(\gamma^\mu
N)_{\gamma}\nonumber\\&+&
\mathcal{A}_4m_{N}^2(\rlap/x\gamma_5C)_{\alpha\beta}N_{\gamma}+
\mathcal{A}_5m_{N}^2(\gamma_\mu\gamma_5
C)_{\alpha\beta}(i\sigma^{\mu\nu}x_\nu N)_{\gamma}+
\mathcal{A}_6m_{N}^3(\rlap/x\gamma_5C)_{\alpha\beta}(\rlap/x
N)_{\gamma}\nonumber\\&+&(\mathcal{T}_1+\frac{x^2m_{N}^2}{4}\mathcal{T}_1^M)(p^\nu
i\sigma_{\mu\nu}C)_{\alpha\beta}(\gamma^\mu\gamma_5
N)_{\gamma}+\mathcal{T}_2m_{N}(x^\mu p^\nu
i\sigma_{\mu\nu}C)_{\alpha\beta}(\gamma_5 N)_{\gamma}\nonumber\\&+&
\mathcal{T}_3m_{N}(\sigma_{\mu\nu}C)_{\alpha\beta}(\sigma^{\mu\nu}\gamma_5
N)_{\gamma}+
\mathcal{T}_4m_{N}(p^\nu\sigma_{\mu\nu}C)_{\alpha\beta}(\sigma^{\mu\rho}x_\rho\gamma_5
N)_{\gamma}\nonumber\\&+& \mathcal{T}_5m_{N}^2(x^\nu
i\sigma_{\mu\nu}C)_{\alpha\beta}(\gamma^\mu\gamma_5 N)_{\gamma}+
\mathcal{T}_6m_{N}^2(x^\mu p^\nu
i\sigma_{\mu\nu}C)_{\alpha\beta}(\rlap/x\gamma_5
N)_{\gamma}\nonumber\\&+&
\mathcal{T}_7m_{N}^2(\sigma_{\mu\nu}C)_{\alpha\beta}(\sigma^{\mu\nu}\rlap/x\gamma_5
N)_{\gamma}+
\mathcal{T}_8m_{N}^3(x^\nu\sigma_{\mu\nu}C)_{\alpha\beta}(\sigma^{\mu\rho}x_\rho\gamma_5
N)_{\gamma} \, \, .
\end{eqnarray}
where the calligraphic functions are functions of the scalar product
($px$) and of the parameters $a_i$, $i=1,~2,~3$. These functions can
be expressed in terms of the nucleon distribution amplitudes with
increasing twist. Explicit expressions of distribution amplitudes
with definite twist are (see also \cite{8,17,18,18p} ):
\begin{eqnarray}\label{sc das}
\mathcal{S}_1 &=& S_1,\nonumber\\
 2px\mathcal{S}_2&=&S_1-S_2,\nonumber\\ \mathcal{P}_1&=&P_1,
\nonumber\\2px\mathcal{P}_2&=&P_1-P_2
\end{eqnarray}

\begin{eqnarray}\label{vec das}
\mathcal{V}_1&=&V_1,\nonumber\\ 2px\mathcal{V}_2&=&V_1-V_2-V_3,
\nonumber\\ 2\mathcal{V}_3&=&V_3,
\nonumber\\ 4px\mathcal{V}_4&=&-2V_1+V_3+V_4+2V_5,\nonumber\\
4px\mathcal{V}_5&=&V_4-V_3,\nonumber\\
4(px)^2\mathcal{V}_6&=&-V_1+V_2+V_3+V_4
 + V_5-V_6
\end{eqnarray}
\begin{eqnarray}\label{ax vec das}
\mathcal{A}_1&=&A_1, \nonumber\\2px\mathcal{A}_2&=&-A_1+A_2-A_3,
\nonumber\\ 2\mathcal{A}_3&=&A_3,
\nonumber\\4px\mathcal{A}_4&=&-2A_1-A_3-A_4+2A_5, \nonumber\\
4px\mathcal{A}_5&=&A_3-A_4\nonumber\\
4(px)^2\mathcal{A}_6&=&A_1-A_2+A_3+A_4-A_5+A_6
\end{eqnarray}
\begin{eqnarray}\label{tens das}
\mathcal{T}_1&=&T_1, \nonumber\\2px\mathcal{T}_2&=&T_1+T_2-2T_3, \nonumber\\
2\mathcal{T}_3&=&T_7,\nonumber\\ 2px\mathcal{T}_4&=&T_1-T_2-2T_7,
\nonumber\\ 2px\mathcal{T}_5&=&-T_1+T_5+2T_8,
\nonumber\\4(px)^2\mathcal{T}_6&=&2T_2-2T_3-2T_4+2T_5+2T_7+2T_8,
\nonumber\\ 4px \mathcal{T}_7&=&T_7-T_8,
\nonumber\\4(px)^2\mathcal{T}_8&=&-T_1+T_2 +T_5-T_6+2T_7+2T_8
\end{eqnarray}
where Eqs. (\ref{sc das}), (\ref{vec das}),  (\ref{ax vec das}) and
 (\ref{tens das}) are for scalar, pseudo scalar, vector, axial
vector and tensor distribution amplitudes respectively. The
distribution amplitudes $F= $ $S_i$, $P_i$, $V_i$, $A_i$, $T_i$ can
be written as:
\begin{equation}\label{dependent1}
F(a_ipx)=\int dx_1dx_2dx_3\delta(x_1+x_2+x_3-1) e^{-ip
x\Sigma_ix_ia_i}F(x_i)\; ,
\end{equation}
where $x_{i}$ with $i=1,~2,~3$ corresponds to the longitudinal
momentum fractions carried by the quarks.

 Using Eqs. (\ref{mut.m})-(\ref{dependent1}) 
 and after carrying out the fourier transformation, the correlator function is obtained in the momentum
 representation. Its explicit expression is given in Appendix B.

 In order to construct sum rules for axial $N-\Delta$ transition
 form factors we need to choose structures. From Eq. (\ref{phen.2})
 it follows that in principle different tensor structures can be
 used for obtaining sum rules for axial form factors. We have
 chosen
 the structures proportional to $g_{\mu\nu}\not\!p'\not\!q$,
 $q_{\mu}p'_{\nu}\not\!p'$, $g_{\mu\nu}\not\!p'$, $q_{\mu}q_{\nu}\not\!p'$
 to obtain sum rules for the form factors $C_{3}^{A}$, $C_{4}^{A}$, $C_{5}^{A}+C_{4}^{A}\frac{p'.q}{m_{\Delta}^{2}}$ and
 $C_{6}^{A}$, respectively. 

 Choosing the coefficients of the structures $g_{\mu\nu}\not\!p'\not\!q$,
 $q_{\mu}p'_{\nu}\not\!p'$, $g_{\mu\nu}\not\!p'$,
 $q_{\mu}q_{\nu}\not\!p'$ in Eqs. (\ref{phen.2}) and (\ref{result1})
 and applying the Borel transformation with respect to the variable
 $p'^{2}=(p-q)^{2}$ which suppress the contributions of the higher
 states and continuum we get desired sum rules for the form factors $C_{3}^{A}$, $C_{4}^{A}$, $C_{5}^{A}$ and
 $C_{6}^{A}$ as: 
\begin{eqnarray}\label{C_{3}}
&&C_{3}(Q^{2})=
\frac{m_{N}}{\sqrt{3}\lambda_{\Delta}} \left. e^{\frac{m_{\Delta}^{2}}{M_{B}^{2}}}\right\{\int_{t_{0}}^{1}dx_{2}\int_{0}^{1-x_{2}}dx_{1}
\frac{e^{-\frac{s(x_{2},Q^{2})}{M_{B}^{2}}}}{x_{2}}\left[2V_{1} -T_{1}\right](x_i)
\nonumber \\ &+&
\int_{t_{0}}^{1}dx_{3}\int_{0}^{1-x_{3}}dx_{1}\frac{e^{-\frac{s(x_{3},Q^{2})}{M_{B}^{2}}}}{x_{3}}T_{1}(x_i')
+\int_{t_0}^1dx_3\int_0^{1-x_3}dx_1\int_{t_0}^{x_3}\frac{dt_1}{t_1^2}e^{-\frac{s(t_1,Q^2)}{M_{B}^{2}
}}\frac{m_{N}^{2}}{M_{B}^{2}}(x_3-t_1){\cal T}_{234578}(x_i')
\nonumber\\
&+&\left.\int_{t_{0}}^{1}dx_{3}\int_{0}^{1-x_{3}}dx_{1}e^{-\frac{s_{0}}{M_{B}^{2}}}\frac{m_{N}^{2}}{Q^{2}+m_{N}^{2}t_{0}^{2}}(-t_{0}+x_{3})
{\cal T}_{234578}(x_i')\right\}
\end{eqnarray}
\begin{eqnarray}\label{C_4}
&&C_{4}(Q^{2}) =\frac{m_N^2}{\sqrt{3}\lambda_\Delta} \left.e^{\frac{m_\Delta^2}{M_B^2}}\right\{
\int_{t_0}^1 dx_2\int_0^{1-x_2}dx_{1}\int_{t_{0}}^{x_{2}}dt_{1}\left. e^{-\frac{s(t_1,Q^2)}{M_B^2}}\right[
\nonumber\\
&+&\frac{2m_{N}}{M_{B}^{2}t_{1}}({\cal V}_{123}-{\cal T}_{123})(x_i)
+\frac{2m_{N}(1-2t_{1})}{M_{B}^{2}t_{1}}(-{\cal A}_{123}+{\cal T}_{127})(x_i)
\nonumber \\ &+&
\frac{4m_{N}^{3}(-1+t_{1})(t_{1}-x_{2})}{M_{B}^{4}t_{1}}{\cal A}_{123456}(x_i)
+\left. \frac{4m_{N}^{3}(-1+t_{1})(t_{1}-x_{2})}{M_{B}^{4}t_{1}}{\cal T}_{125678}(x_i)\right]
\nonumber\\
&+&\int_{t_{0}}^{1}dx_{2}\int_{0}^{1-x_{2}}dx_{1}\left.e^{-\frac{s_{0}}{M_{B}^{2}}}\right[(\frac{2m_{N}t_{0}}{Q^{2}+m_{N}^{2}t_{0}^{2}})
({\cal V}_{123}-{\cal T}_{123})(x_i)
\nonumber \\ &-&
\frac{2m_{N}t_{0}(-1+2t_{0})}{Q^{2}+m_{N}^{2}t_{0}^{2}}(-{\cal A}_{123}+{\cal T}_{127})(x_i)
\nonumber\\
&+&\frac{4 m_N^3 t_0}{Q^2 + m_N^2 t_0^2}
\left\{(-1+t_{0})(-x_{2}+t_{0})\left( \frac{2m_{N}^{2}t_{0}^{5}}{(Q^{2}+m_{N}^{2}t_{0}^{2})^{2}}+\frac{1}{M_B^2} \right) \right.
\nonumber \\
&-& \left. \left. \frac{t_{0}^{2}(4x_{2}-5(1+x_{2}))t_{0}+6t_{0}^{2})}{(Q^{2}+m_{N}^{2}t_{0}^{2})} \right\}
  ({\cal A}_{123456}+{\cal T}_{125678})(x_i)\right]
\nonumber\\
&+&\int_{t_{0}}^{1}dx_{3}\int_{0}^{1-x_{3}}dx_{1}\int_{t_{0}}^{x_{3}}dt_{1}e^{-\frac{s(t_{1},Q^{2})}{M_{B}^{2}}}
\left[-\frac{m_{N}}{M_{B}^{2}}(2{\cal V}_{123}+{\cal A}_{123})(x_i') \right.
-\frac{2m_{N}(-1+t_{1})}{M_{B}^{2}t_{1}}{\cal T}_{127}(x_i')
\nonumber \\
&+&\left.\frac{m_{N}^{3}(-1+t_{1})(t_{1}-x_{3})}{M_{B}^{4}t_{1}}(-2{\cal V}_{123456} - {\cal A}_{123456} + 2{\cal T}_{125678})(x_i')\right]
\nonumber\\&+&
\int_{t_{0}}^{1}dx_{3}\int_{0}^{1-x_{3}}dx_{1}e^{-\frac{s_{0}}{M_{B}^{2}}}\left[- \frac{m_{N}t_{0}^{2}}{Q^{2}+m_{N}^{2}t_{0}^{2}} \right.
(2 {\cal V}_{123}+{\cal A}_{123})(x_i')
-\frac{2m_{N}t_{0}(-1+t_{0})}{Q^{2}+m_{N}^{2}t_{0}^{2}}{\cal T}_{127}(x_i')
\nonumber \\
&+&\left(\frac{2m_{N}^{5}(-1+t_{0})t_{0}^{5}(-x_{3}+t_{0})}{(Q^{2}+m_{N}^{2}t_{0}^{2})^{3}} \right.
-\frac{m_{N}^{3}t_{0}^{3}(4x_{3}-5(1+x_{3})t_{0}+6t_{0}^{2})}{(Q^{2}+m_{N}^{2}t_{0}^{2})^{2}}
\nonumber \\
&+&\left. \left. \left. \frac{m_{N}^{3}(-1+t_{0})t_{0}(-x_{3}+t_{0})}{M_{B}^{2}(Q^{2}+m_{N}^{2}t_{0}^{2} )}\right)
(2{\cal V}_{123456}-{\cal A}_{123456}+2{\cal T}_{125678})(x_i')\right]\right\}
\end{eqnarray}
\begin{eqnarray}\label{C_{5}}
&&C_{5}(Q^{2}) =\frac{1}{\sqrt{3}\lambda_{\Delta}}
\left.e^{\frac{m_{\Delta}^{2}}{M_{B}^{2}}}\right\{\int_{t_{0}}^{1}dx_{2}\int_{0}^{1-x_{2}}dx_{1}e^{-\frac{s(x_{2},Q^{2})}{M_{B}^{2}}}
m_{N}(-S_{1}+P_{1} -V_{3} +A_{3})(x_i)
\nonumber \\
&+&\int_{t_{0}}^{1}dx_{3}\int_{0}^{1-x_{3}}dx_{1}e^{-\frac{s(x_{3},Q^{2})}{M_{B}^{2}}}  m_{N}\left(-V_{3} +\frac{A_{3}}{2}\right)(x_i')
\nonumber \\
&+&\int_{t_{0}}^{1}dx_{2}\int_{0}^{1-x_{2}}dx_{1}\int_{t_{0}}^{x_{2}}dt_{1}e^{-\frac{s(t_{1},Q^{2})}{M_{B}^{2}}} \frac{m_N}{t_1}
\nonumber\\
&\times&\left[\left(2 -\frac{Q^{2}+m_{N}^{2}(-1+2t_{1})+s(t_{1},Q^{2})}{M_{B}^{2}}\right)\right. {\cal V}_{123}(x_i)
+\frac{m_{N}^{2}(t_{1}-x_{2})}{M_{B}^{2}}T_{125678}(x_i)
\nonumber\\
&+&\left. \frac{Q^{2}+m_{N}^{2}(-1+2t_{1})+s(t_{1},Q^{2})}{2M_{B}^{2}}(-2{\cal A}_{123}+{\cal T}_{123} + {\cal T}_{127})(x_i)
\right]
\nonumber\\
&+&\int_{t_{0}}^{1}dx_{2}\int_{0}^{1-x_{2}}dx_{1}\left. e^{-\frac{s_{0}}{M_{B}^{2}}}\frac{m_N t_0}{Q^2+m_N^2 t_0^2} \right[
+m_{N}^{2}t_{0}(t_{0}-x_{2}){\cal T}_{125678}(x_i)
\nonumber \\
&&\left. (-m_{N}^{2}+Q^{2}+s(t_{0},Q^{2})+2m_{N}^{2}t_{0})
(-{\cal V}_{123}-{\cal A}_{123}+{\cal T}_{123}+{\cal T}_{127})(x_i)\right]
\nonumber\\
&+&\int_{t_{0}}^{1}dx_{3}\int_{0}^{1-x_{3}}dx_{1}\int_{t_{0}}^{x_{3}}dt_{1}
e^{-\frac{s(t_{1},Q^{2})}{M_{B}^{2} }}
\frac{m_{N}^{3}(t_{1}-x_{3})}{2M_{B}^{2}t_{1}}
(2{\cal V}_{123456}(x_i')-{\cal A}_{123456}(x_i'))
\nonumber\\
&+&\int_{t_{0}}^{1}dx_{3}\int_{0}^{1-x_{3}}dx_{1}e^{-\frac{s_{0}}{M_{B}^{2}}}\left.
\frac{m_{N}^{3}(t_{0}-x_{3})t_{0}}{2(Q^{2} +M^{2}t_{0}^{2})}
(2{\cal V}_{123456}(x_i')-{\cal A}_{123456}(x_i'))\right\}
\nonumber\\
&-&\frac{(m_{N}^{2}-m_{\Delta}^{2}+Q^{2})}{2m_{N}^{2}}C_{4}(Q^{2},M_{B}^{2})
\end{eqnarray}
\begin{eqnarray}\label{C_{6}}
&&C_{6}(Q^{2}) =\frac{m_{N}^2}{\sqrt{3}\lambda_{\Delta}}
\left. e^{\frac{m_{\Delta}^{2}}{M_{B}^{2}}}\right\{\int_{t_{0}}^{1}dx_{2}\int_{0}^{1-x_{2}}dx_{1}\int_{t_{0}}^{x_{2}}dt_{1}e^{-\frac{s(t_{1},Q^{2})}{M_{B}^{2}
}}
\nonumber\\&&\left[\frac{-2m_{N}(-1+t_{1})}{M_{B}^{2}t_{1}}(-2{\cal A}_{123}+{\cal T}_{123})(x_i)
+\frac{-2m_{N}(-1+t_{1})(-1+2t_{1})}{M_{B}^{2}t_{1}^{2}}{\cal T}_{127}(x_i)\right.
\nonumber\\
&+&\left. \frac{2m_{N}^{3}(2(-1+t_{1})^{2}t_{1}-(1+2(-2+t_{1})t_{1})x_{2})}{M_{B}^{4}t_{1}^{2}}({\cal A}_{123456} + {\cal T}_{125678})(x_i)\right]
\nonumber\\
&+&\int_{t_{0}}^{1}dx_{2}\int_{0}^{1-x_{2}}dx_{1}e^{-\frac{s_{0}}{M_{B}^{2}}}
\left[\frac{4m_{N}(-1+t_{0})t_{0}}{Q^{2}+m_{N}^{2}t_{0}^{2}}{\cal A}_{123}(x_i) \right.
\nonumber\\
&-&2\frac{m_{N}(-1+t_{0})}{Q^{2}+m_{N}^{2}t_{0}^{2}}{\cal T}_{123}(x_i)
-\frac{2m_{N}(-1+t_{0})(-1+2t_{0})}{Q^{2} +m_{N}^{2}t_{0}^{2}}{\cal T}_{127}(x_i)
\nonumber\\
&+&\left(\frac{8m_{N}^{5}(-1+t_{0})^{2}t_{0}^{4}(-x_{2}+t_{0})}{(Q^{2}+m_{N}^{2}t_{0}^{2} )^{3}}\right.
-\frac{4m_{N}^{3}(-1+t_{0})t_{0}^{2}(3x_{2}+t_{0}(-4-5x_{2}+6t_{0}))}{(Q^{2}+m_{N}^{2}t_{0}^{2}
)^{2}}
\nonumber\\
&+&\left. \left. \frac{4m_{N}^{3}(-1+t_{0})^{2}(-x_{2}+t_{0})}{M_{B}^{2}(Q^{2}+m_{N}^{2}t_{0}^{2} )}\right)
({\cal A}_{123456}+{\cal T}_{125678})(x_i)\right]
\nonumber \\
&+&\int_{t_{0}}^{1}dx_{3}\int_{0}^{1-x_{3}}dx_{1}\int_{t_{0}}^{x_{3}}dt_{1}e^{-\frac{s(t_{1},Q^{2})}{M_{B}^{2}
}} \frac{m_N (1-t_1)}{M_B^2 t_1}
\nonumber\\&&
\left[{\cal A}_{123}(x_i')
+\frac{1}{t_1}{\cal T}_{123}(x_i')
+\frac{(-1+2t_{1})}{t_{1}}{\cal T}_{127}(x_i')\right.
\nonumber \\
&+&\left.\frac{2m_{N}^{2}(1-t_{1})(t_{1}-x_{3})}{M_{B}^{2}t_{1}}({\cal V}_{123456}-\frac{1}{2}{\cal A}_{123456}+{\cal T}_{125678})(x_i')\right]
\nonumber \\
&+&\left.\int_{t_{0}}^{1}dx_{3}\int_{0}^{1-x_{3}}dx_{1}e^{-\frac{s_{0}}{M_{B}^{2}}} \frac{m_N (-1+t_0)}{Q^2+m_N^2 t_0^2}
\right[t_{0}{\cal A}_{123}(x_i')+{\cal T}_{123}(x_i')
-(-1+2t_{0}){\cal T}_{127}(x_i')
\nonumber \\
&+&2 m_N^2\left(\frac{2m_{N}^{2}(-1+t_{0})t_{0}^{4}(-x_{3}+t_{0})}{(Q^{2}+m_{N}^{2}t_{0}^{2})^{2}} \right.
-\frac{t_{0}^{2}(3x_{3}+t_{0}(-4-5x_{3}+6t_{0}))}{(Q^{2}+m_{N}^{2}t_{0}^{2})}
\nonumber \\
&+&\left. \left. \left.\frac{(-1+t_{0})(-x_{3}+t_{0})}{M_{B}^{2}}\right)
({\cal V}_{123456}-\frac{1}{2}{\cal A}_{123456}+{\cal T}_{125678})(x_i')\right]\right\}
\end{eqnarray}
where following \cite{18}, the following shorthand notations for various combinations of the DA's are employed:
\begin{eqnarray}
{\cal T}_{234578} &=& T_2 - T_3 - T_4  + T_5  + T_7 + T_8
\nonumber \\
{\cal T}_{123} &=& T_1 + T_2 - 2 T_3
\nonumber \\
{\cal T}_{127} &=&T_1 - T_2 - 2 T_7
\nonumber \\
{\cal T}_{125678} &=& -T_1 + T_2 + T_5 - T_6 + 2 T_7 + 2 T_8
\nonumber \\
{\cal V}_{123} &=& V_1 - V_2 - V_3
\nonumber \\
{\cal V}_{123456} &=& - V_1 + V_2 + V_3 + V_4 + V_5 - V_6
\nonumber \\
{\cal A}_{123} &=& A_1 - A_2 + A_3
\nonumber \\
{\cal A}_{123456} &=& A_1 - A_2 + A_3 + A_4 - A_5 + A_6
\end{eqnarray}
and
\begin{eqnarray}
{\cal F}(x_i) &=& {\cal F}(x_1,x_2,1-x_1-x_2)
\nonumber \\
{\cal F}(x_i') &=& {\cal F}(x_1,1-x_1-x_3,x_3)
\end{eqnarray}
 The Borel transformation and the continuum subtraction are
 performed using the following substitutions (see \cite{17,18}):
\begin{eqnarray}\label{borel}
\int dt \frac{\rho(t)}{(q-tp)^2}&=&\int_0^1 \frac{dt}{t}
\frac{\rho(t)}{s-{p'}^2}\Rightarrow \int_{t_0}^1 \frac{dt}{t}
\rho(t)e^{-\frac{s}{M_B^2}} ,
\nonumber \\
\int dt \frac{\rho(t)}{(q-tp)^4}&=&\int_0^1 \frac{dt}{t^2}
\frac{\rho(t)}{(s-{p'}^2)^2}\Rightarrow  \frac{1}{M_B^2}\int_{t_0}^1
\frac{dt}{t^2}
\rho(t)e^{-\frac{s}{M_B^2}}+\frac{\rho(t_0)e^{-\frac{s_0}{M_B^2}}}{Q^2+t_0^2
m_{N}^2} , \nonumber\\
\int dt \frac{\rho(t)}{(q-tp)^6}&=&\int_0^1 \frac{dt}{t^3}
\frac{\rho(t)}{(s-{p'}^2)^3}\Rightarrow \frac{1}{2M_B^4}\int_{t_0}^1
\frac{dt}{t^3}
\rho(t)e^{-\frac{s}{M_B^2}}\nonumber \\
&&+\frac{\rho(t_0)e^{-\frac{s_0}{M_B^2}}}{2t_0(Q^2+t_0^2
m_{N}^2)M_B^2}-\frac{t_0^2}{2(Q^2+t_0^2m_{N}^2)}\left[\frac{d}{dt_0}\frac{\rho(t_0)}{t_0(Q^2+t_0^2m_{N}^2)}\right] e^{-\frac{s_0}{M_B^2}}, \nonumber\\
 s(t,Q^2)&=&(1-t)M^2+\frac{(1-t)}{t}Q^2, \nonumber\\
t_0(s_0,Q^2)&=&\frac{\sqrt{(Q^2+s_0-m_{N}^2)^2+4m_{N}^2Q^2}-(Q^2+s_0-m_{N}^2)}{2m_{N}^2}.
\end{eqnarray}
where $Q^{2}=-q^{2}$ and $t_{0}$ is the solution of the
$s(t_0,Q^2)=s_{0}$. The terms $\sim e^{-\frac{s_{0}}{M_B^2}}$ are so-called
surface terms which appear in successive partial integrations to
reduce the power of denominators. In the hadronic representation for
the correlator functions the Borel transformation is performed by
substituting $\frac{1}{m_{\Delta}^{2}-p'^{2}}\rightarrow
e^{-\frac{m_{\Delta}^{2}}{M_B^2}}$.
\section{Numerical analysis}
From explicit expressions of the sum rules for the axial $N-\Delta$
transition form factors it follows that the main input parameters
are the nucleon distribution amplitudes (DA's). In general these
distribution amplitudes contain hadronic parameters which should be
determined by some means. Various methods to determine these parameters
give different results. In this work, we considered all three different
determinations of these parameters: a) QCD sum
rules based DA's, where corrections to the DA's are taken into
account and the parameters in DA's are determined from QCD sum
rules, b) A model for nucleon DA's where parameters are chosen in a
such way that the nucleon electromagnetic and axial form factors are
described well within LCSR and c) Asymptotic forms
of DA's of all twists (see for example \cite{17} ).

Explicit expressions of corresponding DA's can be found in
\cite{17} and for completeness we present their expressions in
Appendix. In the appendix, the values of the hadronic parameters obtained
from three different methods is also presented.

For the numerical evaluation of the sum rules for the $N-\Delta$
transition form factors we need also specify the values of the
residue of $\Delta$ baryon $\lambda_{\Delta}$, the continuum
threshold $s_{0}$ and Borel parameter $M_B^{2}$. The residue
$\lambda_{\Delta}$ and $s_{0}$ are determined from analysis of the
mass sum rules: $\lambda_{\Delta}=0.038 ~GeV^{3}$ and $s_{0}=2.6-3
~GeV^{2}$ \cite{19, 20, 21, 22, 23} which we have used in numerical
calculations.

The Borel mass parameter $M_{B}^{2}$ is the auxiliary parameter of
sum rules. Therefore we need to find a suitable region of
$M_{B}^{2}$, where physical results are independent of this
parameter. A suitable region of $M_{B}^{2}$ is determined in
the following way. From one side, $M_{B}^{2}$ has to be small enough in
order to guarantee suppression of higher resonances and the
continuum contributions to the correlation function and from other
side it should be large enough in order to guarantee convergence of
the light cone expansion with increasing twist in QCD calculation.

\begin{figure}[h!]
\begin{center}
\includegraphics[width=9cm]{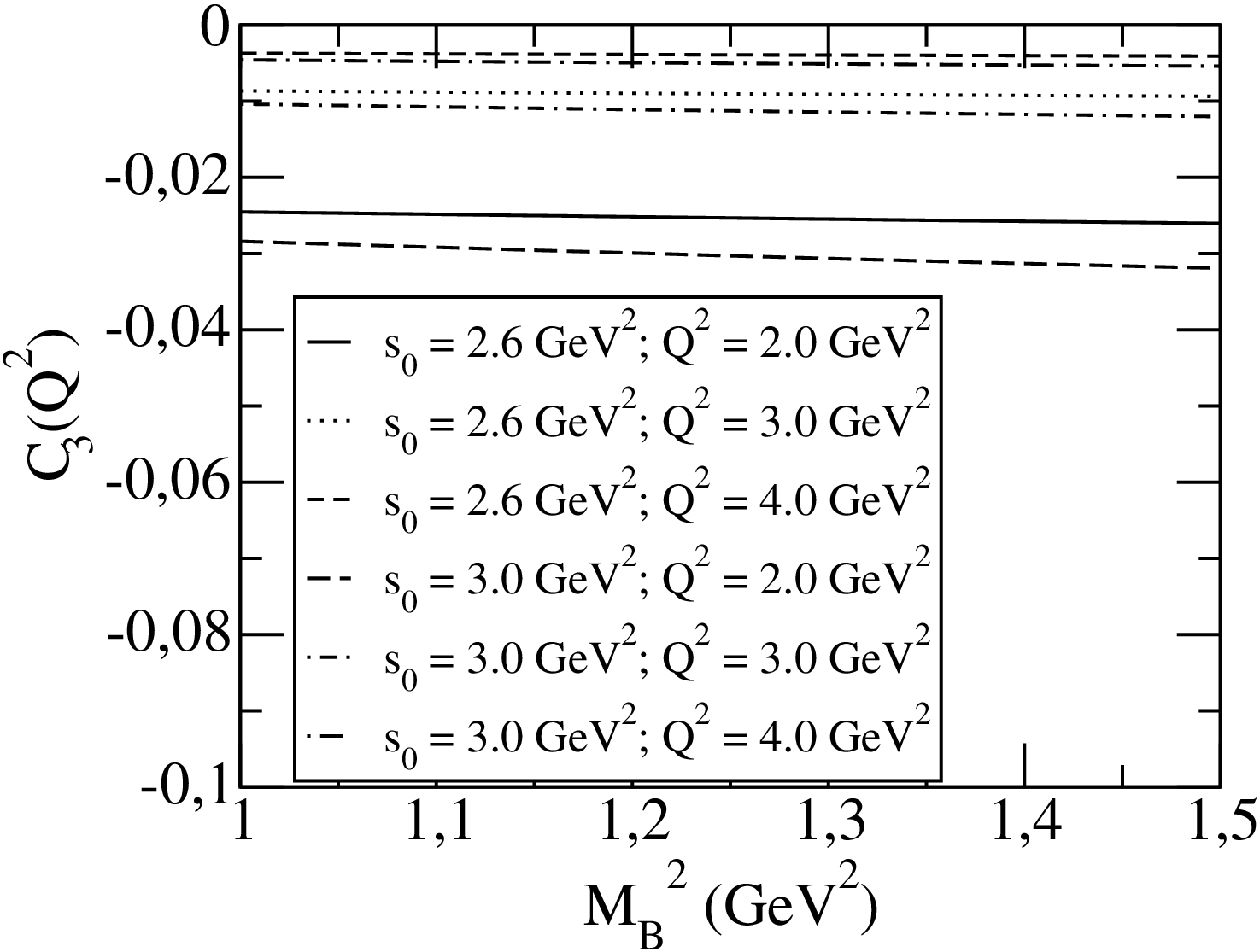}
\end{center}
\caption{The dependence of the form factor $C_3(Q^2)$ on the Borel
parameter squared $M_B^2$ for the values of the continuum threshold
$s_0= 2.6~GeV^2$ and $s_0=3.0~GeV^2$ and at the values of $Q^2 = 3
\pm 1 ~GeV^2$} \label{fig1}
\end{figure}

\begin{figure}[h!]
\begin{center}
\includegraphics[width=9cm]{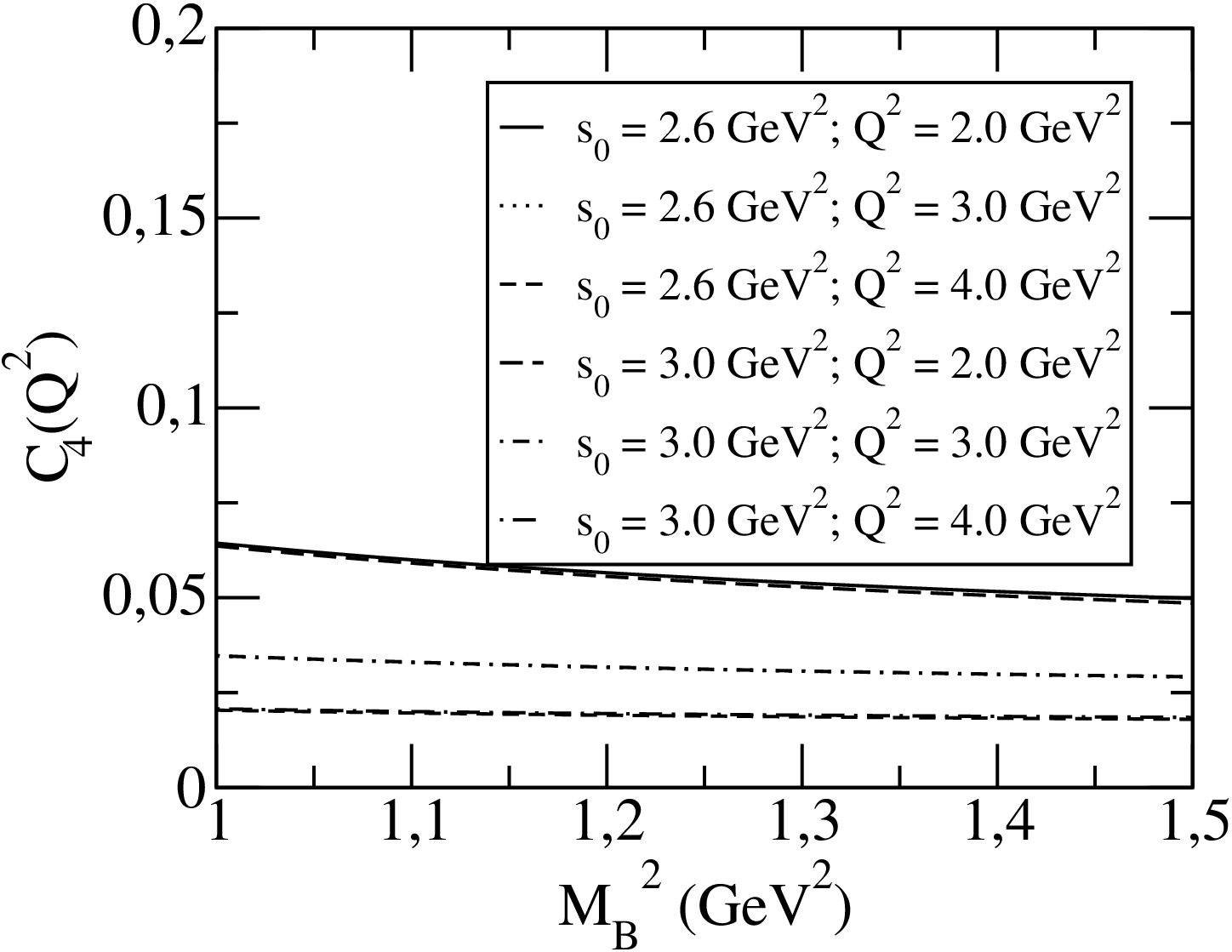}
\end{center}
\caption{The same as Fig. (\ref{fig1}) but for the form factor $C_4(Q^2)$.} \label{fig3}
\end{figure}

\begin{figure}[h!]
\begin{center}
\includegraphics[width=9cm]{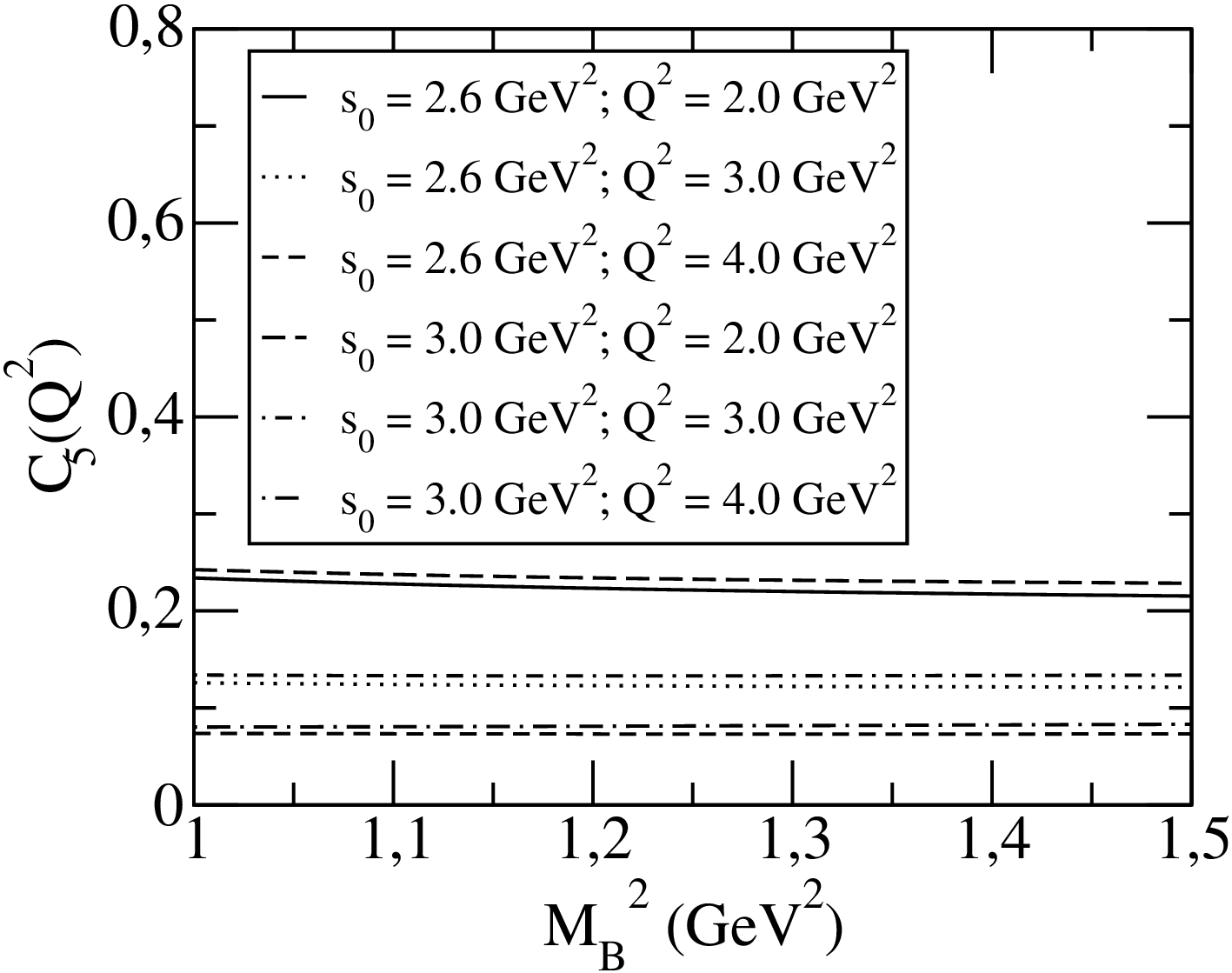}
\end{center}
\caption{The same as Fig. (\ref{fig1}) but for the form factor $C_5(Q^2)$.} \label{fig5}
\end{figure}

\begin{figure}[h!]
\begin{center}
\includegraphics[width=9cm]{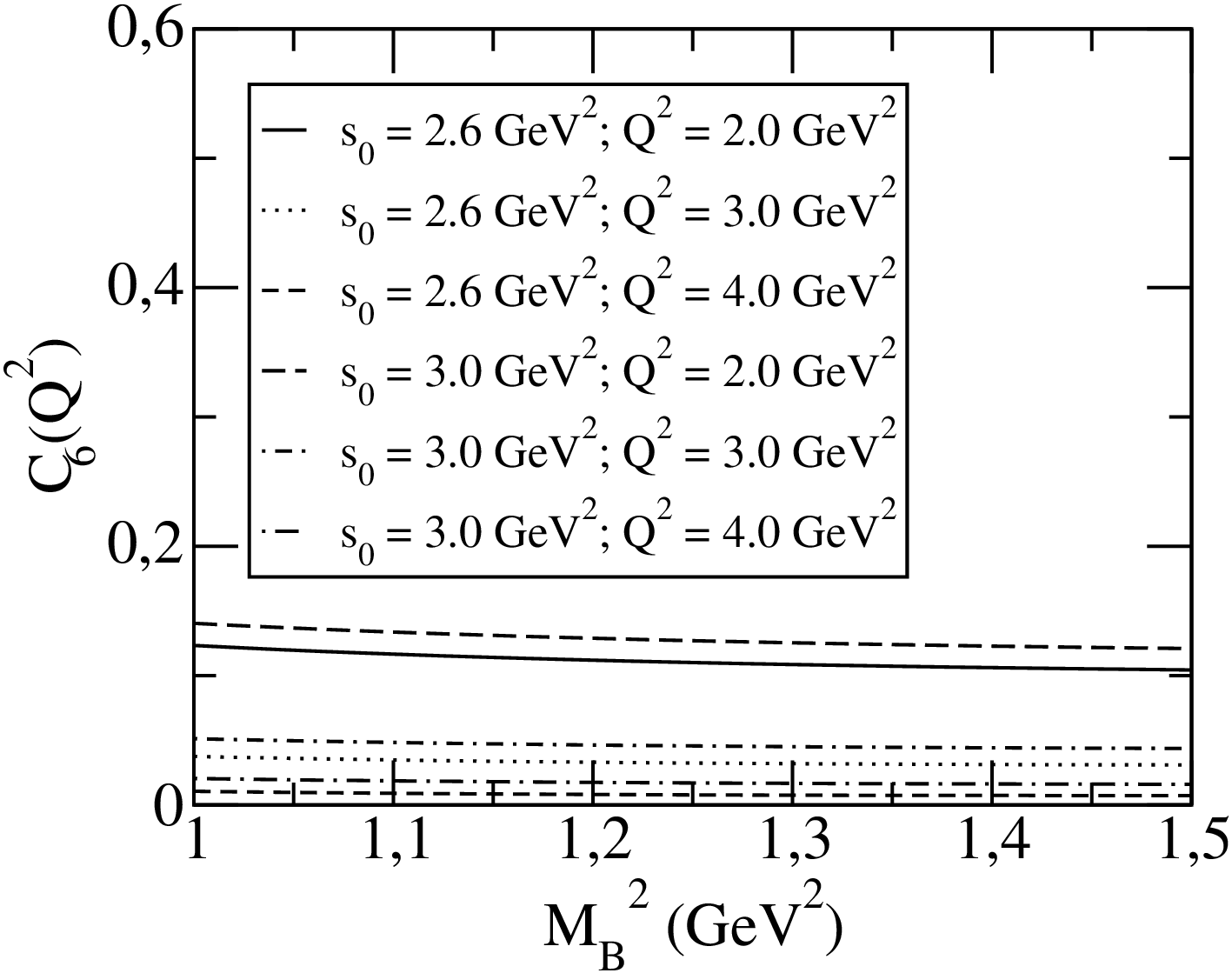}
\end{center}
\caption{The same as Fig. (\ref{fig1}) but for the form factor $C_6(Q^2)$.} \label{fig7}
\end{figure}

In Figs. \ref{fig1}, \ref{fig3}, \ref{fig5}, and
\ref{fig7}, we present the dependence of the form factors
$C_{3}(Q^{2})$, $C_{4}(Q^{2})$, $C_{5}(Q^{2})$ and $C_{6}(Q^{2})$ on
$M_{B}^{2}$ for first set of DA's with two fixed values of $s_{0}$
and three fixed values of  $Q^{2}$ respectively. From these figures
it follows that all considered form factors exhibit good
stability with respect to variation of $M_{B}^{2}$ in the region
$1.2~GeV^{2}\leq M_{B}^{2}\leq 2~GeV^{2}$, so this region of
$M_{B}^{2}$ can be considered as a working region where form factors
are practically independent of $M_{B}^{2}$. We performed similar
analysis for the two other sets of DA's and obtained that within the above
mentioned region of $M_{B}^{2}$,
the form factors are rather stable to variation of $M_{B}^{2}$.

\begin{figure}[h!]
\begin{center}
\includegraphics[width=9cm]{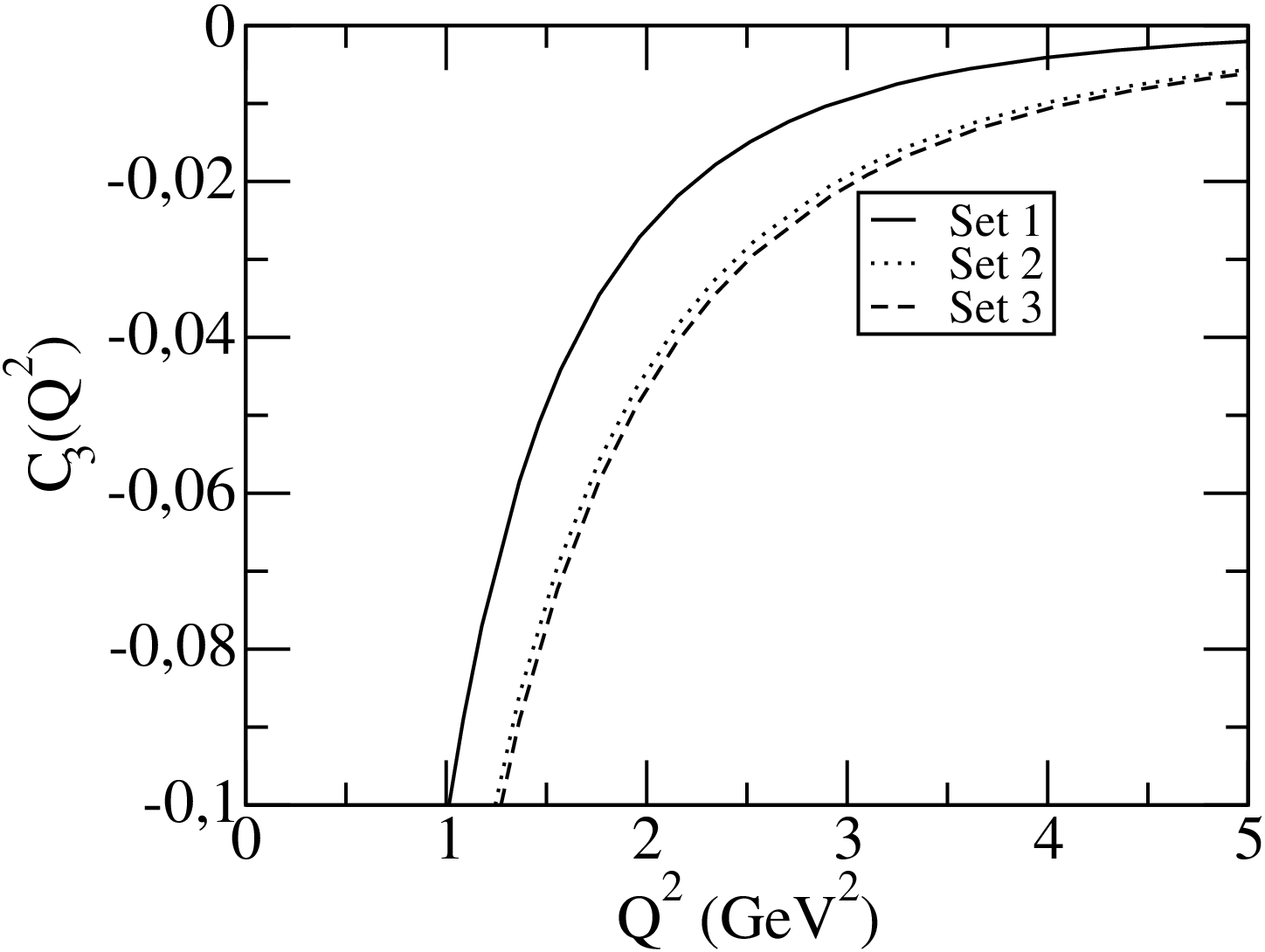}
\end{center}
\caption{The dependence of the form factor $C_3(Q^2)$ for three
different sets of distribution amplitudes at the continuum threshold
$s_{0}=2.6~GeV^{2}$ and the Borel parameter $M_B^{2}=1.5~GeV^{2}$}
\label{fig9}
\end{figure}

\begin{figure}[h!]
\begin{center}
\includegraphics[width=9cm]{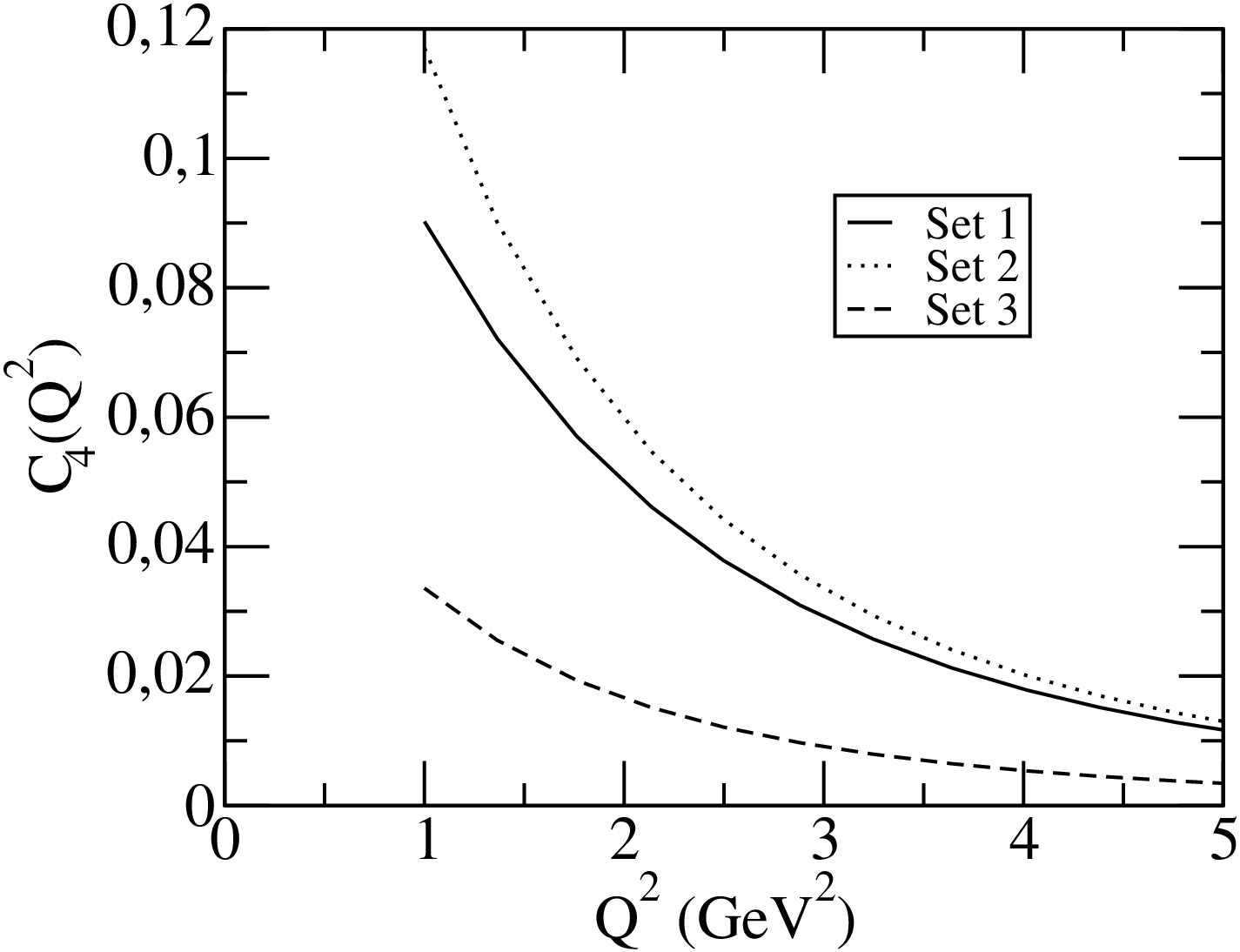}
\end{center}
\caption{The same as Fig. (\ref{fig9}) but for the form factor $C_4(Q^2)$ \hfill } \label{fig10}
\end{figure}

\begin{figure}[h!]
\begin{center}
\includegraphics[width=9cm]{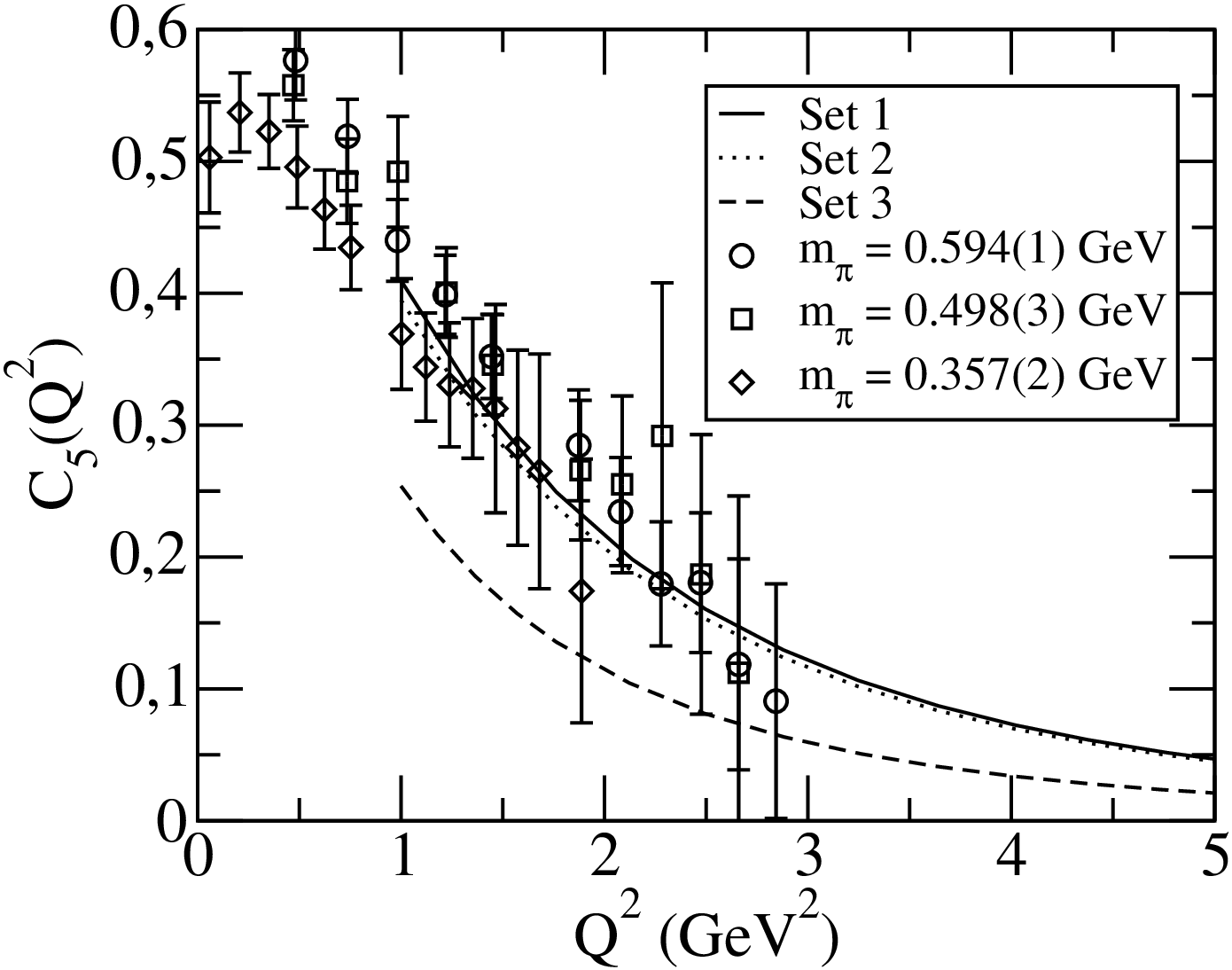}
\end{center}
\caption{The same as Fig. (\ref{fig9}) but for the form factor $C_5(Q^2)$.
Results from the lattice are also shown.} \label{fig11}
\end{figure}

\begin{figure}[h!]
\begin{center}
\includegraphics[width=9cm]{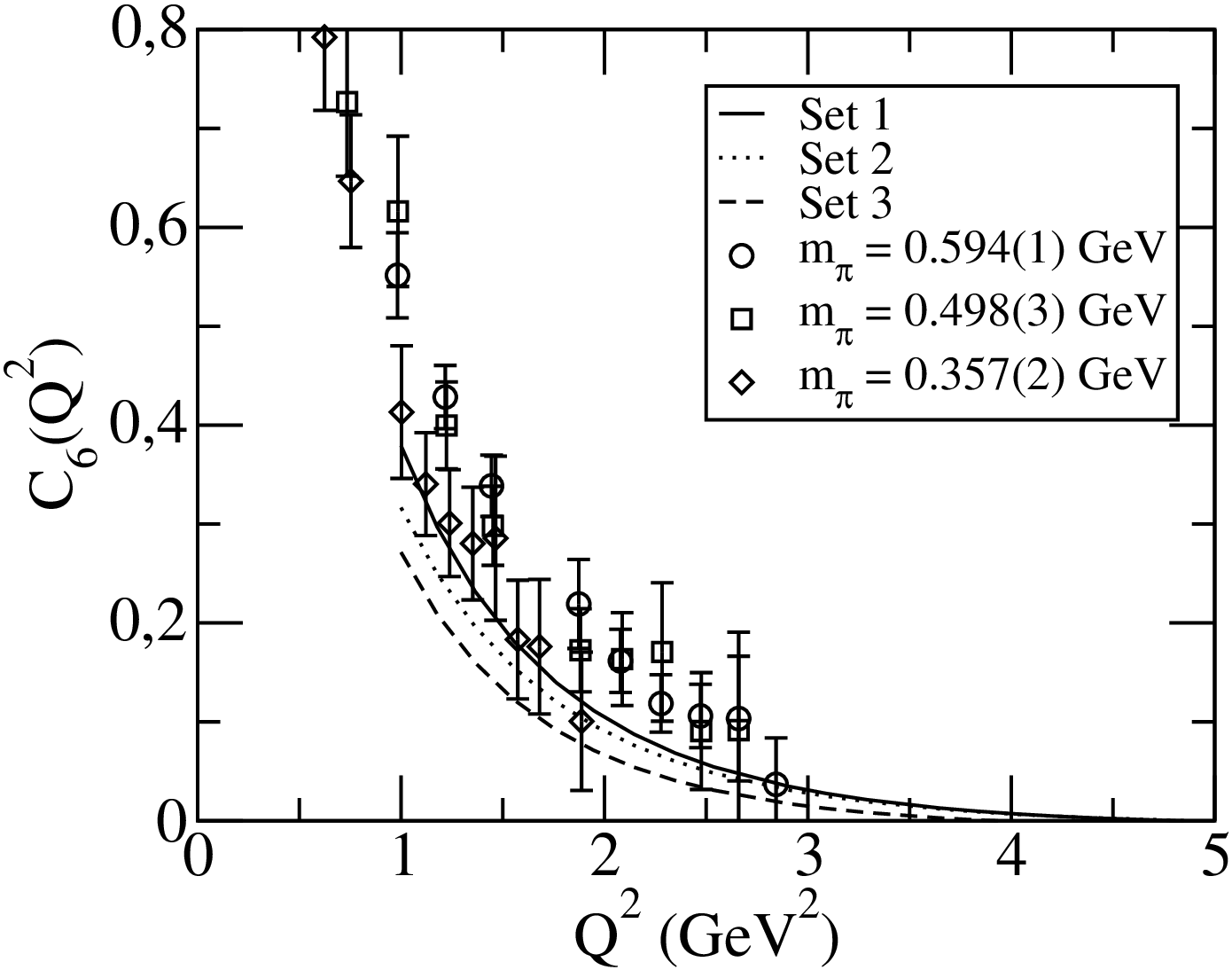}
\end{center}
\caption{The same as Fig. (\ref{fig11}) but for the form factor $C_6(Q^2)$} \label{fig12}
\end{figure}

In Figs. \ref{fig9}, \ref{fig10},  \ref{fig11} and \ref{fig12} we
present the dependence of the form factors $C_i(Q^2)$, $i=3,4,5,6$
on $Q^{2}$ at fixed
values of $M_{B}^{2}$ and $s_{0}$ for all three sets of DA's. From
these figures we see that form factors are very sensitive to the
choice of DA's. For form factor $C_{3}(Q^{2})$ sets 2 and 3 leads to
the same result, for $C_{4}(Q^{2})$ and $C_{5}(Q^{2})$ form factors
sets 1 and 2 give results which are very close to each other and for
$C_{6}(Q^{2})$ up to $Q^{2}=4~GeV^{2} $ all sets of DA's leads to
the different results and when $Q^{2}\geq4~GeV^{2} $ all three sets
of DA's leads to indistinguishable predictions. Our results on form
factors $C_{4}(Q^{2})$ and $C_{5}(Q^{2})$ for three sets of DA's
satisfy the relation $C_{4}(Q^{2})=C_{5}(Q^{2})/4$ assumed in the
experimental analysis.

Figs. (\ref{fig11}) and (\ref{fig12}) contain also the predictions of
lattice calculations from \cite{yeni}. The points correspond to the
values obtained using a hybrid action and assuming $Z_A=1.1$ (see
\cite{yeni} for details of the notation). Note that, due to different
conventions used, the lattice results has been multiplied with $\sqrt{2/3}$.
From the figures, it is seen
that there is a good agreement between the lattice results and our predictions
within error bars. Note also that, the biggest difference between sum rules
predictions and lattice calculations is seen in $C_5(Q^2)$ when the
asymptotic distributions are used, i.e. when the hadronic parameters from
Set 3 are used.

\begin{figure}[h!]
\begin{center}
\includegraphics[width=9cm]{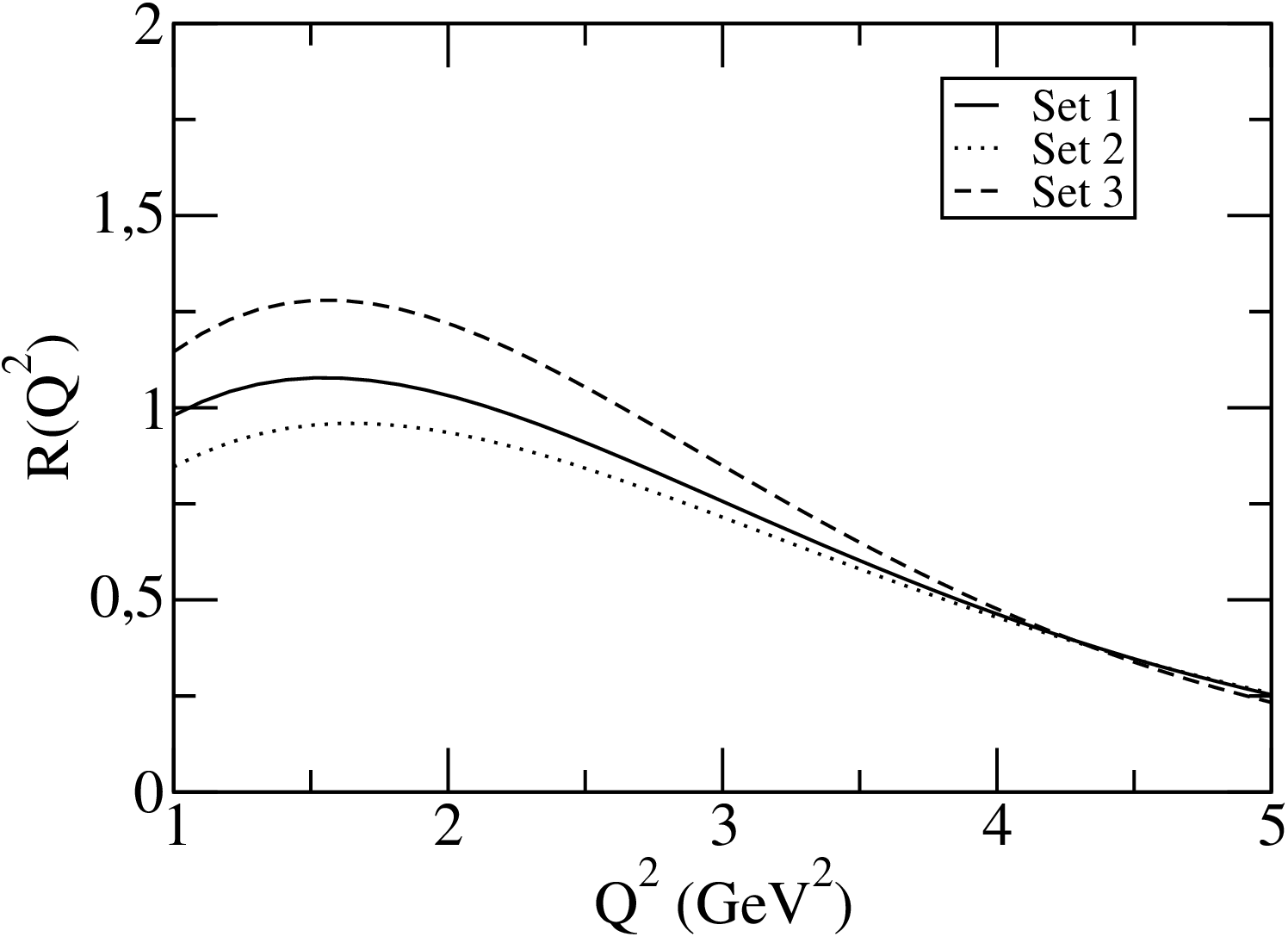}
\end{center}
\caption{The dependence of $R(Q^2)$ as a function of $Q^2$ for the three
sets of DA's.} \label{fig13}
\end{figure}

And finally, in Fig. (\ref{fig13}), $R(Q^2)$ is plotted as a function of
$Q^2$. The function $R(Q^2)$ is defined as:
\begin{eqnarray}
R(Q^2) = \frac{C_6^A(Q^2)}{C_5^A(Q^2)} \frac{Q^2 + m_\pi^2}{m_N^2}
\end{eqnarray}
From Eq. (\ref{reln}), it is seen that, assuming PCAC and pion dominance,
$R(Q^2)=1$. From Fig. (\ref{fig13}), it is seen that at $Q^2 \simeq
1.5~GeV$, PCAC and pion dominance approximations are valid. But for larger
values of $Q^2$, $R(Q^2)$ deviates considerable from unity, and hence we
may conclude that PCAC and pion dominance assumptions break down.


As we have already noted, the axial form factors for the $\Delta \rightarrow N$ transition
have also been calculated in the framework of the constituent quark model in \cite{14}.
Our predictions on these
form factors differ from the results of \cite{14}.

In summary, we have calculated axial $N-\Delta$ transition form
factors which play important role for understanding the dynamics of
weak $N-\Delta$ transition and compare our results with existing
lattice and quark model predictions.
\section{Acknowledgment}
This work is partially supported by TUBITAK under the project number
106T333. One of the authors, A.O., would like to thank TUBA for the
funds provided under the GEBIP program. Also, K. Azizi would like to
thank TUBITAK for their partially support.
 \clearpage
 \begin{appendix}
\section{Appendix A}
\renewcommand{\theequation}{\Alph{section}.\arabic{equation}}

 \setcounter{equation}{0}
In Eqs. (\ref{sc das}), (\ref{vec das}), (\ref{ax vec das}) and
(\ref{tens das}) the distribution amplitudes depend on scale and can
be expanded with the conformal operators, to the next-to-leading
conformal spin accuracy, they are obtained in \cite{17, 18,18p}:
\begin{eqnarray}\label{DA's}
V_1(x_i,\mu)&=&120x_1x_2x_3[\phi_3^0(\mu)+\phi_3^+(\mu)(1-3x_3)],\nonumber\\
V_2(x_i,\mu)&=&24x_1x_2[\phi_4^0(\mu)+\phi_3^+(\mu)(1-5x_3)],\nonumber\\
V_3(x_i,\mu)&=&12x_3\{\psi_4^0(\mu)(1-x_3)+\psi_4^-(\mu)[x_1^2+x_2^2-x_3(1-x_3)]
\nonumber\\&&+\psi_4^+(\mu)(1-x_3-10x_1x_2)\},\nonumber\\
V_4(x_i,\mu)&=&3\{\psi_5^0(\mu)(1-x_3)+\psi_5^-(\mu)[2x_1x_2-x_3(1-x_3)]
\nonumber\\&&+\psi_5^+(\mu)[1-x_3-2(x_1^2+x_2^2)]\},\nonumber\\
V_5(x_i,\mu)&=&6x_3[\phi_5^0(\mu)+\phi_5^+(\mu)(1-2x_3)],\nonumber\\
V_6(x_i,\mu)&=&2[\phi_6^0(\mu)+\phi_6^+(\mu)(1-3x_3)],\nonumber\\
A_1(x_i,\mu)&=&120x_1x_2x_3\phi_3^-(\mu)(x_2-x_1),\nonumber\\
A_2(x_i,\mu)&=&24x_1x_2\phi_4^-(\mu)(x_2-x_1),\nonumber\\
A_3(x_i,\mu)&=&12x_3(x_2-x_1)\{(\psi_4^0(\mu)+\psi_4^+(\mu))+\psi_4^-(\mu)(1-2x_3)
\},\nonumber\\
A_4(x_i,\mu)&=&3(x_2-x_1)\{-\psi_5^0(\mu)+\psi_5^-(\mu)x_3
+\psi_5^+(\mu)(1-2x_3)\},\nonumber\\
A_5(x_i,\mu)&=&6x_3(x_2-x_1)\phi_5^-(\mu)\nonumber\\
A_6(x_i,\mu)&=&2(x_2-x_1)\phi_6^-(\mu),\nonumber\\
T_1(x_i,\mu)&=&120x_1x_2x_3[\phi_3^0(\mu)+\frac{1}{2}(\phi_3^--\phi_3^+)(\mu)(1-3x_3)
],\nonumber\\
T_2(x_i,\mu)&=&24x_1x_2[\xi_4^0(\mu)+\xi_4^+(\mu)(1-5x_3)],\nonumber\\
T_3(x_i,\mu)&=&6x_3\{(\xi_4^0+\phi_4^0+\psi_4^0)(\mu)(1-x_3)+
(\xi_4^-+\phi_4^--\psi_4^-)(\mu)[x_1^2+x_2^2-x_3(1-x_3)]
\nonumber\\
&&+(\xi_4^++\phi_4^++\psi_4^+)(\mu)(1-x_3-10x_1x_2)\},\nonumber\\
T_4(x_i,\mu)&=&\frac{3}{2}\{(\xi_5^0+\phi_5^0+\psi_5^0)(\mu)(1-x_3)+
(\xi_5^-+\phi_5^--\psi_5^-)(\mu)[2x_1x_2-x_3(1-x_3)]
\nonumber\\
&&+(\xi_5^++\phi_5^++\psi_5^+)(\mu)(1-x_3-2(x_1^2+x_2^2))\},\nonumber\\
T_5(x_i,\mu)&=&6x_3[\xi_5^0(\mu)+\xi_5^+(\mu)(1-2x_3)],\nonumber\\
T_6(x_i,\mu)&=&2[\phi_6^0(\mu)+\frac{1}{2}(\phi_6^--\phi_6^+)(\mu)(1-3x_3)],
\nonumber \\
T_7(x_i,\mu)&=&6x_3\{(-\xi_4^0+\phi_4^0+\psi_4^0)(\mu)(1-x_3)+
(-\xi_4^-+\phi_4^--\psi_4^-)(\mu)[x_1^2+x_2^2-x_3(1-x_3)]
\nonumber\\
&&+(-\xi_4^++\phi_4^++\psi_4^+)(\mu)(1-x_3-10x_1x_2)\},\nonumber\\
T_8(x_i,\mu)&=&\frac{3}{2}\{(-\xi_5^0+\phi_5^0+\psi_5^0)(\mu)(1-x_3)+
(-\xi_5^-+\phi_5^--\psi_5^-)(\mu)[2x_1x_2-x_3(1-x_3)]
\nonumber\\
&&+(-\xi_5^++\phi_5^++\psi_5^+)(\mu)(1-x_3-2(x_1^2+x_2^2))\},\nonumber\\
S_1(x_i,\mu) &=& 6 x_3 (x_2-x_1) \left[ (\xi_4^0 + \phi_4^0 +
\psi_4^0 + \xi_4^+ + \phi_4^+ + \psi_4^+)(\mu) + (\xi_4^- + \phi_4^-
- \psi_4^-)(\mu)(1-2 x_3) \right]
\nonumber \\
S_2(x_i,\mu) &=& \frac{3}{2} (x_2 -x_1) \left[- \left(\psi_5^0 +
\phi_5^0 + \xi_5^0\right)(\mu) + \left(\xi_5^- + \phi_5^- - \psi_5^0
\right)(\mu) x_3 \right. \nonumber \\
 && \left.+\left(\xi_5^+ + \phi_5^+ + \psi_5^0 \right)(\mu) (1- 2
x_3)\right]
\nonumber \\
P_1(x_i,\mu) &=& 6 x_3 (x_2-x_1) \left[ (\xi_4^0 - \phi_4^0 -
\psi_4^0 + \xi_4^+ - \phi_4^+ - \psi_4^+)(\mu) + (\xi_4^- - \phi_4^-
+ \psi_4^-)(\mu)(1-2 x_3) \right]
\nonumber \\
P_2(x_i,\mu) &=& \frac32 (x_2 -x_1) \left[\left(\psi_5^0 + \psi_5^0
- \xi_5^0\right)(\mu) + \left(\xi_5^- - \phi_5^- + \psi_5^0
\right)(\mu) x_3 \right. \nonumber\\
&& \left. + \left(\xi_5^+ - \phi_5^+ - \psi_5^0 \right)(\mu) (1- 2
x_3)\right]\, .
\end{eqnarray}
 the parameters used above are defined in
terms of the following eight independent parameters $f_N$,
$\lambda_1$, $\lambda_2$, $V_1^d$, $A_1^u$, $f_d^1$, $f_d^2$ and
$f_u^1$ as
\begin{eqnarray}
\phi_3^0& =& \phi_6^0 = f_N \nonumber \\
 \phi_4^0 &=& \phi_5^0 =
\frac{1}{2} \left(\lambda_1 + f_N\right)
\nonumber \\
\xi_4^0 &=& \xi_5^0 = \frac{1}{6} \lambda_2\nonumber \\
  \psi_4^0
&=& \psi_5^0 = \frac{1}{2}\left(f_N - \lambda_1 \right) \nonumber\\
\phi_3^- &=& \frac{21}{2} A_1^u,\nonumber\\
\phi_3^+ &=& \frac{7}{2} (1 - 3 V_1^d),\nonumber\\
\phi_4^- &=& \frac{5}{4} \left(\lambda_1(1- 2 f_1^d -4 f_1^u) + f_N(
2 A_1^u - 1)\right) \,,
\nonumber \\
\phi_4^+ &=& \frac{1}{4} \left( \lambda_1(3- 10 f_1^d) - f_N( 10
V_1^d - 3)\right)\,,
\nonumber \\
\psi_4^- &=& - \frac{5}{4} \left(\lambda_1(2- 7 f_1^d + f_1^u) +
f_N(A_1^u + 3 V_1^d - 2)\right) \,,
\nonumber \\
\psi_4^+ &=& - \frac{1}{4} \left(\lambda_1 (- 2 + 5 f_1^d + 5 f_1^u)
+ f_N( 2 + 5 A_1^u - 5 V_1^d)\right)\,,
\nonumber \\
\xi_4^- &=& \frac{5}{16} \lambda_2(4- 15 f_2^d)\,,
\nonumber \\
\xi_4^+ &=& \frac{1}{16} \lambda_2 (4- 15 f_2^d)\,,\nonumber\\
\phi_5^- &=& \frac{5}{3} \left(\lambda_1(f_1^d - f_1^u) + f_N( 2
A_1^u - 1)\right) \,,
\nonumber \\
\phi_5^+ &=& - \frac{5}{6} \left(\lambda_1 (4 f_1^d - 1) + f_N( 3 +
4 V_1^d)\right)\,,
\nonumber \\
\psi_5^- &=& \frac{5}{3} \left(\lambda_1 (f_1^d - f_1^u) + f_N( 2 -
A_1^u - 3 V_1^d)\right)\,,
\nonumber \\
\psi_5^+ &=& -\frac{5}{6} \left(\lambda_1 (- 1 + 2 f_1^d +2 f_1^u) +
f_N( 5 + 2 A_1^u -2 V_1^d)\right)\,,
\nonumber \\
\xi_5^- &=& - \frac{5}{4} \lambda_2 f_2^d\,,
\nonumber \\
\xi_5^+ &=&  \frac{5}{36} \lambda_2 (2 - 9 f_2^d)\,,
\nonumber \\
\phi_6^- &=& \phantom{-}\frac{1}{2} \left(\lambda_1 (1- 4 f_1^d - 2
f_1^u) + f_N(1 +  4 A_1^u )\right) \,,
\nonumber \\
\phi_6^+ &=& - \frac{1}{2}\left(\lambda_1  (1 - 2 f_1^d) + f_N ( 4
V_1^d - 1)\right)
\end{eqnarray}
Our numerical values are obtained using:
\begin{eqnarray}
f_{N} &=& (5.0\pm0.5)\times10^{-3}~GeV^{2},
\nonumber \\
\lambda_{1} &=& -(2.7\pm0.9)\times10^{-2}~GeV^{2}, \nonumber\\
\lambda_{2} &=& (5.4\pm1.9)\times10^{-2}~GeV^{2}
\end{eqnarray}
And for other five independent parameter we have used three sets as:
\begin{eqnarray}
\mbox{Set 1 }\nonumber\\
 A_{1}^{u} &=& 0.38\pm0.15,\nonumber \\
V_{1}^{d} &=& 0.23\pm0.03, \nonumber\\
f_{1}^{d} &=& 0.40\pm0.05, \nonumber \\
f_{2}^{d} &=& 0.22\pm0.05,\nonumber \\
f_{1}^{u} &=& 0.07\pm0.05
\\
\mbox{Set 2 }\nonumber\\
 A_{1}^{u} &=& \frac{1}{14},\nonumber \\
V_{1}^{d} &=& \frac{13}{42}, \nonumber\\
f_{1}^{d} &=& 0.40\pm0.05, \nonumber \\
f_{2}^{d} &=& 0.22\pm0.05,\nonumber \\
f_{1}^{u} &=& 0.07\pm0.05
\end{eqnarray}
\begin{eqnarray}
\mbox{Set 3 }\nonumber\\
 A_{1}^{u} &= 0,\nonumber \\
V_{1}^{d} &= \frac{1}{3}, \nonumber\\
f_{1}^{d} &= \frac{3}{10}, \nonumber \\
f_{2}^{d} &= \frac{4}{15},\nonumber \\
f_{1}^{u} &= \frac{1}{10}
\end{eqnarray}
Note that the asymptotic forms of DA's can be obtained from
Eqs.(\ref{DA's}) using the values given in the $3^{rd}$ set.

\section{Appendix B}
In this appendix, we present the explicit form of the correlation function containing all the Dirac structures.
\begin{eqnarray}\label{result1}
&&\Pi_{\mu\nu}(p,q)=
\left. \left. \frac{i}{\sqrt{3}}\right\{\int_{0}^{1}dx_{1}\int_{0}^{1-x_{1}}dx_{2}\frac{1}{(q-x_{2}p)^{2}} \right[
\nonumber\\
&&m_{N}\left[\not\!p'g_{\mu\nu} x_{2}
+(\gamma_{\nu}q_\mu-\not\!qg_{\mu\nu})(1-x_{2})\right](S_{1}-P_{1}+V_{3}-A_{3})(x_i)
\nonumber \\
&+&2\left[-\not\!p'\not\!qg_{\mu\nu}+\not\!p'\gamma_{\nu}q_{\mu}+g_{\mu\nu}p'.q-p'_{\nu}q_{\mu}\right]V_{1}(x_i)
\nonumber\\
&+&\left[2q_{\mu}(p'_{\nu}+2q_{\nu})(-1+x_{2})+g_{\mu\nu}\left(p'.q+p.(q-x_{2}p)+q^{2}-(p'+q)^2\right)x_{2}+2p'_{\nu}q_{\mu}x_{2}\right]A_{1}(x_i)
\nonumber \\
&+&\left\{\not\!p'\not\!qg_{\mu\nu}+\gamma_{\nu}\not\!qq_{\mu}(-1+x_{2})-\not\!p'\gamma_{\nu}q_{\mu}(1+x_{2})-g_{\mu\nu}x_2 (2p'.q+q^{2}) \right.
\nonumber\\
&-&\left.  \vphantom{\int_0^{x_3}}\left. 2q_{\mu}(p'_{\nu}+2q_{\nu})(-1+x_{2})x_2+g_{\mu\nu}(p'+q)^{2}x_{2}\right\}T_{1}(x_{i})\right]
\nonumber\\
&+& \left. \int_{0}^{1}dx_{1}\int_{0}^{1-x_{1}}dx_{3}\frac{1}{(q-x_{3}p)^{2}}\right[
\nonumber \\
&& m_{N}\left[\not\!p'g_{\mu\nu}x_{3}+(\gamma_{\nu}q_\mu-\not\!qg_{\mu\nu}) (1-x_{3})\right]\left(V_{3}-\frac{A_{3}}{2}\right)(x_i')
\nonumber \\
&+&q_{\mu}\left[(\gamma_{\nu}\not\!q-2q_{\nu})(-1+x_{3})-\not\!p\gamma_{\nu} x_{3}\right]\left(V_{1}+\frac{A_{1}}{2}\right)(x_i')
\nonumber \\
&+&\left.\left[(\not\!p'\gamma_{\nu}+2q_{\nu})q_{\mu}-g_{\mu\nu}(\not\!p'\not\!q +q^{2})
-2q_{\mu}(p'_{\nu}+q_{\nu})x_3 +g_{\mu\nu}(p'+q)^{2}x_{3}\right]T_{1}(x_i')\right]
\nonumber\\
&+&\left. \left. \int_{0}^{1}dx_{1}\int_{0}^{1-x_{1}}dx_{2}\int_{0}^{x_{2}}dt_{1}\right(\frac{1}{(q-t_{1}p)^{6}}\right[
\nonumber \\
&-&8m_{N}^3(-1+t_{1})(t_{1}-x_{2})q_{\mu}\left[t_{1}p'_{\nu}+(-1+t_{1})q_{\nu}\right] \left[t_{1}\not\!p'+(-1+t_{1})\not\!q\right]{\cal A}_{123456}(x_i)
\nonumber \\
&+&4m_{N}^{2}(-1+t_{1})(t_{1}-x_{2})q_{\mu}
\nonumber \\
&\times& \left\{p.(q-t_{1}p)\left[t_{1}\not\!p'\gamma_{\nu}-(-1+t_{1})\gamma_{\nu}\not\!q\right] \right.
\nonumber \\
&&+t_{1}p'_{\nu}\left[2\not\!p'\not\!q+p.(q-t_{1}p)+t_{1}p'^{2}+(-3+2t_{1})p'.q+(-1+t_{1})q^{2}\right]
\nonumber \\
&& \left.  -(1-t_{1})q_{\nu}\left[2\not\!p'\not\!q+3p.(q-t_{1}p)+t_{1}p'^{2}+(-3+2t_{1})p'.q+(-1+t_{1})q^{2}\right]\right\}{\cal T}_{234578}(x_i')
\nonumber \\
&-& \left. \vphantom{\int_0^{x_3}} 8m_{N}^{3}(-1+t_{1})(t_{1}-x_{2})q_{\mu}\left[t_{1}p'_{\nu}-(1-t_{1})q_{\nu}\right]\left[t_{1}\not\!p'-(1-t_{1})\not\!q\right]{\cal T}_{125678}(x_i')\right]
\nonumber \\
&+&\frac{1}{(q-t_{1}p)^{4}}\left[ \vphantom{\int_0^1}2m_{N}^{3}(1-t_{1})(t_{1}-x_{2})q_{\mu}\gamma_{\nu}{\cal A}_{123456}(x_i) \right.
\nonumber \\
&+&m_{N}^{2}(t_{1}-x_{2})\left[g_{\mu\nu}\left(p.(q-t_{1}p)+t_{1}p'^{2}+(-1+2t_{1})p'.q+(-1+t_{1})q^{2}\right) \right.
\nonumber \\
&&+\left. q_{\mu}\left\{(3-2t_{1})\not\!p'\gamma_{\nu}+2(-1+t_{1})\left(\gamma_{\nu}\not\!q-2 q_\nu\right)-2p'_{\nu}\vphantom{x^2}\right\}\right]{\cal T}_{234578}(x_i)
\nonumber \\
&-&m_{N}^{3}(t_{1}-x_{2})\left[g_{\mu\nu}\not\!p'+(-1+t_{1})(g_{\mu\nu}\not\!q+q_{\mu}\gamma_{\nu})\right]{\cal T}_{125678}(x_i)
\nonumber \\
&+&m_{N}^{2}q_{\mu}(-1+t_{1})\left[\gamma_{\nu}\not\!q(-1+t_{1})-\not\!p'\gamma_{\nu}t_{1}+2p'_{\nu}t_{1}\right](S_{1}-S_{2}+P_{1}-P_{2})(x_i)
\nonumber \\
&+&m_{N}\left[2\not\!p'\gamma_{\nu}\not\!qq_{\mu}-2\not\!q(g_{\mu\nu}p.(q-t_{1}p)-p'_{\nu} q_{\mu})(-1+t_{1}) \vphantom{x^2}\right.
\nonumber \\
&&-2\left(\rule{0pt}{10pt}\not\!p'g_{\mu\nu}p.(q-t_{1}p)+\not\!p'\gamma_{\nu}\not\!qq_{\mu}+\not\!p'p'_{\nu}q_{\mu}\right)t_{1}
\nonumber \\
&&\left. +\gamma_{\nu}q_{\mu}(-1+t_{1})(-3p'.q+3p.(q-t_{1}p)-q^{2}+(p'+q)^{2}t_{1})\right]{\cal V}_{123}(x_i)
\nonumber\\
&-&\frac{1}{2}m_{N}^{2}\left\{\vphantom{\frac{}{}}q_{\mu}(1-t_{1})\left[3\gamma_{\nu}\not\!q(1-t_{1})+\not\!q\gamma_{\nu}(1-t_{1})+2\not\!p'\gamma_{\nu}t_{1}\right] \right.
\nonumber \\
&&\left. +\left[-2q_{\mu}(1-t_{1})\left(q_{\nu}(1-t_{1})+2p'_{\nu}t_{1}\right)+g_{\mu\nu}\left\{q^{2}-2(p'.q+q^{2})t_{1}+(p'+q)^{2}t_{1}^{2}\right\}\right]\vphantom{\frac{}{}}\right\}(V_{4}-V_{3})(x_i)
\nonumber \\
&-&m_{N}\left[\vphantom{x^2}\left\{2\not\!q(-1+t_{1})+2\not\!p't_{1}\right\}\left\{g_{\mu\nu}p.(q-t_{1}p)-q_{\mu}(p'_{\nu}+2q_{\nu})+2q_{\mu}(p'_{\nu} +q_{\nu})t_{1}\right\} \right.
\nonumber \\
&&- \left. \gamma_{\nu}q_{\mu}(-1+t_{1})\left\{-p'.q+p.(q-t_{1}p)-q^{2}+(p'+q)^2t_{1}\right\}\right]{\cal A}_{123}(x_i)
\nonumber \\
&-&2q_{\mu}M^{2}(-1+t_{1})\left[q_{\nu}(-1+t_{1})+p'_{\nu}t_{1}\right]{\cal A}_{1345}(x_i)
\nonumber \\
&-&q_{\mu}M^{2}(-1+t_{1})\left[\left(\gamma_{\nu}\not\!q-2 q_\nu \right)(-1+t_{1})-\not\!p'\gamma_{\nu}t_{1}\right](A_{3}-A_{4})(x_i)
\nonumber\\
&+&m_{N}\left[\vphantom{\frac{}{}}(\not\!qg_{\mu\nu}-\gamma_\nu q_\mu) p.(q-t_{1}p)(-1+t_{1})\right.
\nonumber \\
&&+\left. \not\!p'\left(-2q_{\mu}q_{\nu} +g_{\mu\nu}p.(q-t_{1}p)t_{1}+2q_{\mu}(p'_{\nu}+q_{\nu})t_{1}\vphantom{x^2}\right)\right]{\cal T}_{123}(x_i)
\nonumber \\
&+&M\left[ \vphantom{\frac{}{}}-\not\!p'\gamma_{\nu}\not\!qq_{\mu}(1-t_1)+2\not\!p'q_{\mu}q_{\nu} +
\not\!p'\left( \vphantom{x^2} g_{\mu\nu}p.(q-t_{1}p)-2q_{\mu}(p'_{\nu}+3q_{\nu})\right)t_{1} \right.
\nonumber \\
&&+4\not\!p'q_{\mu}(p'_{\nu} +q_{\nu})t_{1}^{2}+\not\!q(-1+t_{1})\left(\vphantom{x^2} g_{\mu\nu}p.(q-t_{1}p)-2q_{\mu}(p'_{\nu}+2q_{\nu})+4q_{\mu}(p'_{\nu}+q_{\nu})t_{1}\right)\nonumber\\
&&-\left. \gamma_{\nu}q_{\mu}(-1+t_{1})\left(\vphantom{x^2} p.(q-t_{1}p)-q^{2}+2p'.q+(p'-q)^{2}t_{1}\right)\vphantom{\frac{}{}} \right]{\cal T}_{127}(x_i)
\nonumber \\
&+&\left. q_{\mu}m_{N}^{2}(-1+t_{1})\left[\vphantom{x^2}(\gamma_{\nu}\not\!q-4 q_\nu)(-1+t_{1})-\not\!p'\gamma_{\nu}t_{1}-2p'_{\nu}t_{1}\right]{\cal T}_{158}(x_i)\vphantom{\int_0^1}\right]
\nonumber \\
&+&\left. \frac{m_N}{2(q-t_{1}p)^{2}}\right[(-(\not\!p'+\not\!q)g_{\mu\nu}-\gamma_{\nu}q_{\mu})(-2{\cal A}_{123}+{\cal T}_{123} + {\cal T}_{127})(x_i)
\nonumber \\
&+&\left.3m_N g_{\mu\nu}(V_{4}-V_{3})(x_i)-2((\not\!p'+ \not\!q)g_{\mu\nu}-\gamma_{\nu}q_{\mu}){\cal V}_{123}(x_i)\vphantom{\frac{1}{x^2}}\right]
\nonumber\\
&+&\left. \left. \int_{0}^{1}dx_{1}\int_{0}^{1-x_{1}}dx_{3}\int_{0}^{x_{3}}dt_{1}\right(\frac{1}{(q-t_{1}p)^{6}}\right[
\nonumber \\
&& 2q_{\mu}m_{N}^{2}(1-t_{1})\left[q_{\nu}-(p'_{\nu}+q_\nu)t_{1}\right]
\left\{\vphantom{x^2} 2\left[\not\!p'\not\!q-p'.q+(p'+q).(q-t_{1}p)\right]{\cal T}_{234578}(x_i') \right.
\nonumber\\
&+&\left. \vphantom{\int_0^1} \left. m_{N}\left[\not\!q-(\not\!p'+\not\!q) t_{1}\right]\left(2{\cal V}_{123456}-{\cal A}_{123456}+2{\cal T}_{125678}\right)(x_i')\vphantom{x^2}\right\}(t_{1}-x_{3})\right]
\nonumber \\
&+&\frac{1}{(q-t_{1}p)^{4}}\left[\frac{1}{2}m_{N}^{3}g_{\mu\nu}(\not\!q(-1+t_{1})+\not\!p't_{1})(t_{1}-x_{3}){\cal A}_{123456}(x_i')\vphantom{\int_0^1}\right.
\nonumber \\
&+&m_{N}^{2}(t_{1}-x_{3})\left[g_{\mu\nu}\left(\vphantom{x^2}\not\!p'\not\!q-p'.q+(p'+q).(q-t_{1}p)\right) \right.
\nonumber \\
&&-\left. q_{\mu}(\gamma_{\nu}\not\!q-2 q_\nu)(1-t_1)-q_{\mu}\gamma_{\nu}\not\!p't_{1}\vphantom{x^2}\right]{\cal T}_{234578}(x_i')
- \vphantom{\int_0^1}m_{N}^{3}(t_{1}-x_{3})(-1+t_{1})q_{\mu}\gamma_{\nu}{\cal T}_{125678}(x_i')
\nonumber \\
&+& 2m_{N}q_{\mu}(\not\!q(-1+t_{1})+ \not\!p't_{1})(q_{\nu}(-1+t_{1})+p'_{\nu}t_{1}){\cal V}_{123}(x_i')
\nonumber \\
&+&\frac{1}{2}q_{\mu}m_{N}^{2}(-1+t_{1})\left[(\not\!q\gamma_{\nu}+2q_{\nu})(-1+t_{1})-\not\!p'\gamma_{\nu}t_{1}\vphantom{x^2}\right]{\cal V}_{1345}(x_i')
\nonumber \\
&-&\frac{1}{2}q_{\mu}m_{N}^{2}(-1+t_{1})\left[(\not\!q\gamma_{\nu}+2q_{\nu})(-1+t_{1})-(\not\!p'\gamma_{\nu}-4p'_{\nu})t_{1}\vphantom{x^2}\right](V_{4}-V_{3})(x_i')
\nonumber \\
&-&g_{\mu\nu}m_{N}^{3}(\not\!q(-1+t_{1})+\not\!p't_{1})(t_{1}-x_{3}){\cal V}_{123456}(x_i')
\nonumber \\
&+& m_{N}q_{\mu}\left[\not\!q(-1+t_{1})+ \not\!p't_{1}\right]\left[q_{\nu}(-1+t_{1})+p'_{\nu}t_{1}\right]{\cal A}_{123}(x_i')
\nonumber \\
&-&\frac{1}{4}q_{\mu}m_{N}^{2}(-1+t_{1})\left(\vphantom{x^2}(\gamma_{\nu}\not\!q-2q_{\nu})(-1+t_{1})-\not\!p'\gamma_{\nu}t_{1}\right){\cal A}_{1345}(x_i')
\nonumber \\
&-&\frac{1}{4}q_{\mu}m_{N}^{2}(-1+t_{1})\left(\vphantom{x^2}(\gamma_{\nu}\not\!q+2q_{\nu})(-1+t_{1})-(\not\!p'\gamma_{\nu}-4p'_{\nu})t_{1}\right)(A_{3}-A_{4})(x_i')
\nonumber\\
&-&\frac{1}{2}m_{N}q_{\mu}(-1+t_{1})\left(\not\!p'\not\!q-2\not\!p'q_{\nu}+\gamma_{\nu}\left[\vphantom{x^2}p'.q-(p'+q).(q-t_{1}p)\right]\right){\cal T}_{123}(x_i')
\nonumber\\
&+&\frac{1}{2}m_{N}q_{\mu}(-1+t_{1})\left(\vphantom{\frac12}-\not\!p'\gamma_{\nu}\not\!q-2(\not\!p'+2\not\!q)q_{\nu}+\gamma_{\nu}\left[\vphantom{x^2}-p'.q+(p'+q).(q-t_{1}p)\right] \right.
\nonumber \\
&&+\left. 4(\not\!p'+\not\!q)(p'_{\nu}+q_{\nu})t_{1}\vphantom{\frac12}\right){\cal T}_{127}(x_i')
-\left. \left. 2q_{\mu}m_{N}^{2}(-1+t_{1})(q_{\nu}(-1+t_{1})+p'_{\nu}t_{1}){\cal T}_{158}(x_i')\left. \vphantom{\frac12}\right]\vphantom{\int_0^1} \right)\right\}N(p)
\nonumber \\
\end{eqnarray}
where $p'=p-q$. The short hand notations for the functions ${\cal A}_X$, ${\cal T}_X$ and ${\cal V}_X$ are defined within the main body of the
text. The shorthand notations that do not appear within the main text are defined as:
\begin{eqnarray}
{\cal A}_{1345} &=& 2 A_1 + A_3 + A_4 - 2 A_5 \nonumber \\
{\cal T}_{158} &=& - T_1 + T_5 + 2 T_8 \nonumber \\
{\cal V}_{1345} &=& - 2 V_1 + V_3 + V_4 + 2 V_5
\end{eqnarray}

\end{appendix}

\newpage
\listoffigures

%
%
%
%
%
%
%
\clearpage
\newpage
\end{document}